\documentstyle[12pt]{article}

\def\journal#1, #2, #3, #4 { {\sl #1~}{\bf #2~} (#3)  #4 }

\def\cmp{\journal Comm. Math. Phys., }

\def\np{\journal Nucl. Phys., }

\def\pl{\journal Phys. Lett., }

\def\ijmp{\journal Int. J. Mod. Phys., }

\catcode`\@=11
\def\marginnote#1{}
\newcount\hour
\newcount\minute
\newtoks\amorpm
\hour=\time\divide\hour by60
\minute=\time{\multiply\hour by60 \global\advance\minute
by-\hour}\edef\standardtime{{\ifnum\hour<12
\global\amorpm={am}%
        \else\global\amorpm={pm}\advance\hour by-12 \fi
        \ifnum\hour=0 \hour=12 \fi
        \number\hour:\ifnum\minute<10
0\fi\number\minute\the\amorpm}}
\edef\militarytime{\number\hour:\ifnum\minute<10
0\fi\number\minute}

\def\draftlabel#1{{\@bsphack\if@filesw {\let\thepage\relax
   \xdef\@gtempa{\write\@auxout{\string
      \newlabel{#1}{{\@currentlabel}{\thepage}}}}}\@gtempa
   \if@nobreak \ifvmode\nobreak\fi\fi\fi\@esphack}
        \gdef\@eqnlabel{#1}}
\def\@eqnlabel{}
\def\@vacuum{}
\def\draftmarginnote#1{\marginpar{\raggedright\scriptsize\tt#1}}
\def\draft{\oddsidemargin -.5truein
        \def\@oddfoot{\sl preliminary draft \hfil
        \rm\thepage\hfil\sl\today\quad\militarytime}
        \let\@evenfoot\@oddfoot \overfullrule 3pt
        \let\label=\draftlabel
        \let\marginnote=\draftmarginnote

\def\@eqnnum{(\theequation)\rlap{\kern\marginparsep\tt\@eqnlabel}%
\global\let\@eqnlabel\@vacuum}  }

\def\numberbysection{\@addtoreset{equation}{section}
        \def\theequation{\thesection.\arabic{equation}}}

\def\underline#1{\relax\ifmmode\@@underline#1\else
 $\@@underline{\hbox{#1}}$\relax\fi}

\catcode`@=12
\relax
\numberbysection

\def\beq{\begin{equation}}
\def\eeq{\end{equation}}
\def\beqa{\begin{eqnarray}}
\def\eeqa{\end{eqnarray}}
 \def\nnn{\nonumber \\}

\def\hhat{{\widehat h}}
\def\qhat{{widehat q}}
\def\lfloorhat{{\hat \lfloor}}
\def\rfloorhat{{\hat \rfloor}}

\def\Jhat{{\widehat J}}
\def\Je{J^e{}}
\def\Jehat{{\Jhat^e}{}}

\def\kappahat{{\widehat \kappa}}
\def\Nhat{{\widehat N}}

\def\Nhat{{\widehat N}}

\def\Mhat{{\widehat M}}

\def\Ms{ M^\circ}
\def\Mshat{{\widehat  M^\circ}}
\def\Msone{M_1^\circ}
\def\Mstwo{M_2^\circ}
\def\Msonehat{\widehat{M_1^\circ}}
\def\Mstwohat{\widehat{M_2^\circ}}

\def\xihat{{\widehat \xi}}

\def\qhat{{\widehat q}}

\def\phat{{\widehat p}}

\def\Jgen#1 {  {\underline J_{#1}} }
\def\Jgenp#1 #2 {(J_{#1}+{#2},\Jhat_{#1})}
\def\Jgenm#1 #2 {(J_{#1}-{#2},\Jhat_{#1})}
\def\Jg#1 {J_{#1},\Jhat_{#1}}
\def\Jgp#1 #2 {J_{#1}+{#2},\Jhat_{#1}}
\def\Mgen#1 {{\underline M_{#1}}}
\def\mgen#1 {{\underline m_{#1}}}

\def\fin{\end{document}}

\def\Jge{{\underline J}}

\def\Jgen#1 {  {\underline J_{#1}} }
\def\Jgenp#1 #2 {(J_{#1}+{#2},\Jhat_{#1})}
\def\Jgenm#1 #2 {(J_{#1}-{#2},\Jhat_{#1})}
\def\Jg#1 {J_{#1},\Jhat_{#1}}
\def\Jgp#1 #2 {J_{#1}+{#2},\Jhat_{#1}}
\def\Mgen#1 {{\underline M_{#1}}}

\def\fusV#1,#2,#3,#4,#5,#6 {f_V(
\Jgen{#1} ,
\Jgen{#2} ,
\Jgen{#3} ,
\Jgen{#4} ,
\Jgen{#5} ,
\Jgen{#6} )}

\def\demi{{{1\over 2}}}
\def\xii{\xi^{({1\over 2})}}

\def\psii{\psi^{({1\over 2})}}

\def\Phat{\widehat P}
\def\Jpo{{\cal O}\! \left[J_+\right]}
\def\Jmo{{\cal O}\! \left[J_-\right]}
\def\Jpmo{{\cal O}\! \left[J_\pm \right]}
\def\Jto{{\cal O}\! \! \left[q^{J_3}\right]}
\def\Jso{{\cal O}\! \! \left[q^{-J_3}\right]}
\def\Jpophat#1{{\cal O}\! \left[\Jhat_+\right]_{#1}}

\def\Jtophat#1{{\cal O}\! \! \left[\qhat^{\Jhat_3}\right]_{\! #1}}

\def\Jtpsi#1,#2,#3,#4,#5{[ q^{{\cal J}_3, #2}_{\phantom{{\cal J}_3}
#5}{}]^{(#1)}_{#3#4}}
\def\Jspsi#1,#2,#3,#4,#5{[ q^{-{\cal J}_3, #2}_{\phantom{-{\cal J}_3}
#5}{}]^{(#1)}_{#3#4}}
\def\Jppsi#1,#2,#3,#4,#5{[ {\cal J}^{\phantom{+,}#2}_{+,
#5}{}]^{(#1)}_{#3#4}}
\def\Jmpsi#1,#2,#3,#4,#5{[ {\cal J}^{\phantom{-,}#2}_{-,
#5}{}]^{(#1)}_{#3#4}}

\def\U#1,#2,#3,#4{{\cal U}^{(#1)}_{#2 #3} (#4)}
\def\V#1,#2,#3,#4{{\cal V}^{(#1)}_{#2 #3} (#4)}

\def\Ms{M^\circ{}}
\def\Ps{P^\circ{}}

\def\Msone{M_1^\circ}
\def\Mstwo{M_2^\circ}
\def\Psone{P_1^\circ}
\def\Pstwo{P_2^\circ}

\def\Mshat{ {\Mhat^\circ}{}}
\def\Pshat{ {\Phat^\circ}{}}

\def\Msonehat{{\Mhat_1^\circ}{}}
\def\Mstwohat{{\Mhat_2^\circ}{}}
\def\Mssum{M_{12}^\circ{}}
\def\Pssum{P_{12}^\circ{}}

\def\Je{J^e}
\def\Ju{{\underline J}}

\def\Musone{{\underline \Msone}}
\def\Mustwo{{\underline \Mstwo}}
\def\Mussum{{\underline \Mssum}}
\def\Pusone{{\underline \Psone}}
\def\Pustwo{{\underline \Pstwo}}
\def\Pussum{{\underline \Pssum}}
\def\Juhat{\widehat{\underline J}}
\def\qJ{q^{J_3}}
\def\qmJ{q^{-J_3}}
\def\qJu{\underline \qJ}
\def\qmJu{\underline \qmJ}
\def\qJhat{\qhat^{\Jhat_3}}
\def\qmJhat{\qhat^{-\Jhat_3}}
\def\qJuhat{\underline \qJhat}
\def\qmJuhat{\underline \qmJhat}
\def\demi{{{1\over 2}}}
\def\xii{\xi^{({1\over 2})}}
\def\xic{\xi^{[\demi,\demi](0)}_0}

\def\psii{\psi^{({1\over 2})}}

\def\Phat{\widehat P}
\def\Jpo{{\cal O}\! \left[J_+\right]}
\def\Jmo{{\cal O}\! \left[J_-\right]}
\def\Jpmo{{\cal O}\! \left[J_\pm \right]}
\def\Jto{{\cal O}\! \! \left[q^{J_3}\right]}
\def\Jso{{\cal O}\! \! \left[q^{-J_3}\right]}
\def\Jtso{{\cal O}\! \! \left[q^{\pm J_3}\right]}

\def\Do{{\cal O}\! \! \left[D\right]}
\def\Jt{{\widetilde J}}
\def\Jpop#1{{\cal O}\! \left[J_+\right]_{#1}^{(R)}}
\def\Jmop#1{{\cal O}\! \left[J_-\right]_{#1}^{(R)}}

\def\Jtop#1{{\cal O}\! \! \left[q^{J_3}\right]_{\! #1}^{(R)}}
\def\Jsop#1{{\cal O}\! \! \left[q^{-J_3}\right]_{\! #1}^{(R)}}
\def\Jpophat#1{{\cal O}\! \left[\Jhat_+\right]_{#1}^{(R)}}

\def\Jtophat#1{{\cal O}\! \! \left[\qhat^{\Jhat_3}\right]_{\!
#1}^{(R)}}

\def\Dop#1,#2{{\cal O}\! \left[D\right]_{#1,#2}^{(R)}}
\def\Jpopl#1{{\cal O}\! \left[J_+\right]_{#1}^{(L)}}
\def\Jmopl#1{{\cal O}\! \left[J_-\right]_{#1}^{(L)}}

\def\Jsopl#1{{\cal O}\! \! \left[q^{-J_3}\right]_{\! #1}^{(L)}}

\def\Top#1,#2{{\cal O}\! \left[T^{#1}\right]_{#2}}
\def\Ttop#1{{\cal O}\! \left[q^{T_3}\right]}
\def\Tsop#1{{\cal O}\! \left[q^{-T_3}\right]}
\def\Ttsop#1{{\cal O}\! \left[q^{\pm T_3}\right]}

\def\Jtpsi#1,#2,#3,#4,#5{[ q^{{\cal J}_3, #2}_{\phantom{{\cal
J}_3}
#5}{}]^{(#1)}_{#3#4}}
\def\Jttpsi#1,#2,#3,#4,#5{[ q^{{2\cal J}_3, #2}_{\phantom{{\cal
J}_3}
#5}{}]^{(#1)}_{#3#4}}
\def\Jspsi#1,#2,#3,#4,#5{[ q^{-{\cal J}_3, #2}_{\phantom{-{\cal
J}_3}
#5}{}]^{(#1)}_{#3#4}}
\def\Jsspsi#1,#2,#3,#4,#5{[ q^{-{2
\cal J}_3, #2}_{\phantom{-{\cal J}_3}
#5}{}]^{(#1)}_{#3#4}}
\def\Jppsi#1,#2,#3,#4,#5{[ {\cal J}^{\phantom{+,}#2}_{+,
#5}{}]^{(#1)}_{#3#4}}
\def\Jpmpsi#1,#2,#3,#4,#5{[ {\cal J}^{\phantom{+,}#2}_{\pm ,
#5}{}]^{(#1)}_{#3#4}}
\def\Xppsi#1,#2,#3,#4,#5{[ {\cal X}^{\phantom{+,}#2}_{+,
#5}{}]^{(#1)}_{#3#4}}
\def\Xpmpsi#1,#2,#3,#4,#5{[ {\cal X}^{\phantom{+,}#2}_{\pm ,
#5}{}]^{(#1)}_{#3#4}}
\def\Jmpsi#1,#2,#3,#4,#5{[ {\cal J}^{\phantom{-,}#2}_{-,
#5}{}]^{(#1)}_{#3#4}}
\def\Xmpsi#1,#2,#3,#4,#5{[ {\cal X}^{\phantom{-,}#2}_{-,
#5}{}]^{(#1)}_{#3#4}}
\def\Tpsi#1,#2,#3,#4,#5{[ {\cal T}^{#2}_{
#5}{}]^{(#1)}_{#3#4}}
\def\LJtpsi#1,#2,#3,#4,#5,#6,#7,#8{
[ \Lambda\left(q^{{\cal J}_3, #3}\right)_{
#8}{}]^{(#1,#2)}_{#4#5\,#6#7}}
\def\LJspsi#1,#2,#3,#4,#5,#6,#7,#8{
[ \Lambda\left(q^{-{\cal J}_3, #3}\right)_{
#8}{}]^{(#1,#2)}_{#4#5\,#6#7}}
\def\LJppsi#1,#2,#3,#4,#5,#6,#7,#8{
[ \Lambda\left({\cal J_+}\right)^{#3}_{
#8}{}]^{(#1,#2)}_{#4#5\,#6#7}}
\def\LJmpsi#1,#2,#3,#4,#5,#6,#7,#8{
[ \Lambda\left({\cal J_-}\right)^{#3}_{
#8}{}]^{(#1,#2)}_{#4#5\,#6#7}}
\def\LJpmpsi#1,#2,#3,#4,#5,#6,#7,#8{
[ \Lambda\left({\cal J_\pm}\right)^{#3}_{
#8}{}]^{(#1,#2)}_{#4#5\,#6#7}}
\def\LXppsi#1,#2,#3,#4,#5,#6,#7,#8{
[ \Lambda\left({\cal  X_+}\right)^{#3}_{
#8}{}]^{(#1,#2)}_{#4#5\,#6#7}}
\def\LXmpsi#1,#2,#3,#4,#5,#6,#7,#8{
[ \Lambda\left({\cal X_-}\right)^{#3}_{
#8}{}]^{(#1,#2)}_{#4#5\,#6#7}}
\def\LXpmpsi#1,#2,#3,#4,#5,#6,#7,#8{
[ \Lambda\left({\cal  X_\pm}\right)^{#3}_{
#8}{}]^{(#1,#2)}_{#4#5\,#6#7}}
\def\LTpsi#1,#2,#3,#4,#5,#6,#7,#8{
[ \Lambda\left({\cal T}\right)^{#3}_{
#8}{}]^{(#1,#2)}_{#4#5\,#6#7}}

\def\U#1,#2,#3,#4{{\cal U}^{(#1)}_{#2 #3} (#4)}
\def\V#1,#2,#3,#4{{\cal V}^{(#1)}_{#2 #3} (#4)}

\def\Ms{M^\circ{}}
\def\Ps{P^\circ{}}

\def\Msone{M_1^\circ}
\def\Mstwo{M_2^\circ}
\def\Psone{P_1^\circ}
\def\Pstwo{P_2^\circ}

\def\Mshat{ {\Mhat^\circ}{}}
\def\Pshat{ {\Phat^\circ}{}}

\def\Msonehat{{\Mhat_1^\circ}{}}
\def\Mstwohat{{\Mhat_2^\circ}{}}
\def\Mssum{M_{12}^\circ{}}
\def\Pssum{P_{12}^\circ{}}

\def\Je{J^e}
\def\Ju{{\underline J}}

\def\Musone{{\underline \Msone}}
\def\Mustwo{{\underline \Mstwo}}
\def\Mussum{{\underline \Mssum}}
\def\Pusone{{\underline \Psone}}
\def\Pustwo{{\underline \Pstwo}}
\def\Pussum{{\underline \Pssum}}
\def\Juhat{\widehat{\underline J}}
\def\qJ{q^{J_3}}
\def\qmJ{q^{-J_3}}
\def\qJu{\underline \qJ}
\def\qmJu{\underline \qmJ}
\def\qJhat{\qhat^{\Jhat_3}}
\def\qmJhat{\qhat^{-\Jhat_3}}
\def\qJuhat{\underline \qJhat}
\def\qmJuhat{\underline \qmJhat}
\begin{document}
\begin{titlepage}

\nopagebreak \begin{flushright}
CERN-TH/95-48,
LPTENS--95/11\\
hep-th@xxx/9503198
 \\
\end{flushright}

\vglue .5   true cm
\begin{center}
{\large \bf
OPERATOR COPRODUCT-REALIZATION  \\
\medskip
OF QUANTUM GROUP TRANSFORMATIONS\\
\medskip
IN TWO DIMENSIONAL GRAVITY I.\footnote{partially supported by the E.U.
network ``Capital Humain et Mobilit\'e'' contracts \#
CHRXCT920069, and CHRXCT920035} }
\vglue .5 true cm
{\bf Eug\`ene CREMMER}\\
\medskip
{\bf Jean-Loup~GERVAIS\footnote{Part time visitor at Theory Division
CERN.}}\\
\medskip
{\footnotesize Laboratoire de Physique Th\'eorique de
l'\'Ecole Normale Sup\'erieure\footnote{Unit\'e Propre du
Centre National de la Recherche Scientifique,
associ\'ee \`a l'\'Ecole Normale Sup\'erieure et \`a
l'Universit\'e
de Paris-Sud.},\\
24 rue Lhomond, 75231 Paris CEDEX 05, ~France.}\\
{\bf Jens SCHNITTGER}\\
{\footnotesize  CERN Theory Division, CH-1211 Geneva 23, Switzerland.
}\\
\medskip
\end{center}
\vglue .5 true cm
\begin{abstract}
\baselineskip .4 true cm
{\footnotesize
\noindent
  A simple  connection between the universal $R$ matrix of
$U_q(sl(2))$  (for spins $\demi$
 and   $J$) and the required form of the co-product
action of the Hilbert space generators
of the quantum group symmetry is put forward.
This leads us to an explicit operator
realization of the co-product action on the covariant operators.
 It allows us to derive the expected quantum
group
covariance of the fusion and braiding matrices,
although it is  of  a new type:
the generators  depend upon worldsheet variables, and  obey a
new central extension of the
$U_q(sl(2))$ algebra realized by (what we call) fixed point commutation
relations. This is explained by showing on a general  ground  that the link
between the algebra of field transformations and that of the
co-product generators is much weaker than previously thought.
The central charges  of our extended $U_q(sl(2))$ algebra, which
includes the Liouville zero-mode momentum in a nontrivial way,
are related to Virasoro-descendants of unity.
 We also show how our approach   can be used  to derive
the Hopf algebra structure of the extended quantum-group symmetry
$U_q(sl(2))\odot U_{\qhat}(sl(2))$ related to the presence
of both of the screening charges of 2D gravity.

}
\end{abstract}
\vfill
\end{titlepage}
\section{Introduction}
The quantum group structure of two dimensional gravity in the conformal gauge
has led to striking developments\cite{B} --\cite{GS3},
by  allowing us to derive general
formulae for the fusion and braiding coefficients of the operator product
algebra (OPA)  in terms of quantum group symbols of $U_q(sl(2))$. Moreover,
 there exists\cite{B,G1,G3,CGR2}  a
covariant basis of holomorphic
operators, where there is a natural quantum group action
which is a symmetry of the OPA. However, this characterization of the quantum
group symmetry
is somewhat implicit, as so far we do not have an explicit construction
of the $U_q(sl(2))$ generators as operators on the Hilbert space of states,
i.e. a
Hamiltonian realization of the quantum group symmetry.
We would like to stress here that the quantum group symmetry we are talking
about is distinct
from the socalled "dressing symmetries"\cite{BeBa}. The latter transform
solutions of the equations of motion into different ones, while here  the
"physical
observables" (the functionals of the Liouville field) are invariant, and the
symmetry
is seen only in the enlarged phase space defined by the free field construction
of the
 Liouville field.
In fact the situation is very similar  for general conformal field theories
related to a
Coulomb-gas
construction, the most basic example being given by  the $c<1$
minimal models, and thus the relevance of this question extends much beyond the
specific
framework within which we will address it to have a definite setting. Gomez and
Sierra\cite{GS}
showed that there exists
a realization of the Borel subalgebra generated by $J_+$ and $J_3$ for the
$c<1$
theories\footnote{This work was generalized to the WZNW theories in
\cite{RRR}.}
in terms of contour-creating operators acting on suitable screened
vertex operators. Though the operator product and the braiding of the latter
do reproduce the q-Clebsch-Gordan coefficients and braiding matrix of
$U_q(sl(2))$,
they are not, strictly speaking, conformal objects. The basic
message from the work of refs.\cite{B,G1,G3,CGR2} is that in fact there
exists a basis of primary conformal fields which have the desired properties.
Furthermore, in analogy to the Gomez-Sierra treatment, the action
 of $J_+$ within the new basis is related to multiplication by a suitably
defined screening
charge\cite{GS2}, though the contour integral realization
of this multiplication (if it exists) is not obvious, and probably in any case
not very
natural. However, as was pointed out in \cite{MaSc},
the  general realization
of quantum group generators, independent of any Coulomb gas picture,
should be given in terms of certain operators on the Hilbert space
acting on covariant fields by braiding, generalizing the action of
"classical" symmetries by commutators. In the present article, we undertake
steps towards a concrete realization of this form. Let us first summarize
the basic point of the general exposition contained in \cite{MaSc}, where
the principles of (quasi) Hopf quantum group symmetry in quantum theory
have been nicely formulated\footnote{(see also \cite{BL} for a general
discussion with
emphasis on  affine quantum group and Yangian symmetry.)}.
First consider a field theory with  an ordinary (not q
deformed) Lie algebra $\cal G$ of symmetries. For any element $J^a\in \cal
G$, there exists  an operator ${\cal O}(J^a)$ such that a typical field
$\Psi_\ell$
transforms as
\beq
\Bigl [ {\cal O}(J^a), \, \Psi_\ell \Bigr ]=
\sum_m  \Psi_m \left[ J^a\right]_{m\ell}.
\label{ord1}
\eeq
In this equation $\left[ J^a\right]_{m \ell}$ is the matrix of the particular
representation of $\cal G$ under which $\Psi_\ell$ transforms. By
text-book calculations, one of course verifies that the group law is
satisfied since
$$
\Bigl [ {\cal O}(J^b), \,\Bigl [ {\cal O}(J^a),\, \Psi_\ell \Bigr ] \Bigr ]=
\sum_{n m}  \Psi_n \left[J^b\right]_{nm}\, \left[J^a\right]_{m\ell},
$$
together with the Jacobi identity which implies
\beq
\Bigl [ \Bigl[ {\cal O}(J^a), \, {\cal O}(J^b)\Bigr],  \, \Psi_\ell  \Bigr ] =
\sum_{n }  \Psi_n \left[J^a, J^b \right]_{n\ell}.
\label{com0}
\eeq
Products  of fields obey
\beq
\Bigl [ {\cal O}(J^a), \, \Psi_{\ell_1} \Psi_{\ell_1} \Bigr ]=
\sum_{m_1 m_2}
\Psi_{m_1}\Psi_{m_2} \Bigl (\left[J^a\right]_{m_1 \ell_1} \delta_{m_2\,
\ell_2} + \delta_{m_1\,
\ell_1} \left[J^a\right]_{m_2 \ell_2}\Bigr),
\label{ord2}
\eeq
so that they transform by a tensor product of representations as
expected. This last point clearly shows  that such generators cannot
exist for quantum groups for which  simple tensor products  of
representations do not form representations. In order to  introduce  the
remedy, let us rewrite Eq.\ref{ord1} under the form
\beq
{\cal O}(J^a) \Psi_\ell=
\Psi_m \Bigl( \left[J^a\right]_{m\ell}\,  {\cal  I}
+ \delta_{m,\, \ell} {\cal O}(J^a)\Bigr)
\label{ord3}
\eeq
where we introduced the identity operator $\cal  I$ in the Hilbert space
of states.   As is well known, standard Lie groups may be regarded as
particular cases  of Hopf algebras  endowed with a  (trivial) coproduct,
namely given an element $a\in {\cal G}$, one lets\footnote{\label{f1} We denote
the
coproducts by the letter $\Lambda$ instead of the more usual letter $\Delta$,
since the latter is used for conformal weights.}
$\Lambda_0(J^a)=J^a\otimes 1+1\otimes J^a$. Clearly, this coproduct appears
in formulae Eqs.\ref{ord2} and \ref{ord3}, the first with the
coproduct between two matrix representations, the second with the
coproduct between a matrix representation and the operator realization
within the Hilbert space of states.

{}From there, the generalization to quantum Hopf algebras becomes
natural, following ref.\cite{MaSc}.
Consider a  quantum deformation of an enveloping algebra
with generators $J^a$, and co-product (recall footnote \ref{f1})
\beq
\Lambda(J^a):=\sum_{cd}\Lambda^a_{cd}J^c\otimes J^d
\label{copgen}
\eeq
which is co-associative
\beq
\sum_e\Lambda_{bc}^e\Lambda_{ed}^a=\sum_e\Lambda_{cd}^e\Lambda_{be}^a.
\label{coassoc}
\eeq
 Then
Eq.\ref{ord3} is to be replaced by
\beq
{\cal O}(J^a) \Psi_\ell=\sum_{m_1,\, m_2} \sum_{b,\, c}
\Psi_m \Lambda_{bc}^a \left[J^b\right]_{m\ell} {\cal O}(J^c).
\label{gen1}
\eeq
The present action is now
consistent. Indeed,  an easy calculation using the
co-associativity shows that  Eq.\ref{ord2} is replaced by
\beq
{\cal O}(J^a) \, \Psi_{\ell_1} \Psi_{\ell_2}=
\Psi_{m_1}\Psi_{m_2} \Lambda_{bc}^a\left\{
\Lambda_{de}^b \left[J^d\right]_{m_1 \ell_1}
\left[J^e\right]_{m_2 \ell_2}\right\} {\cal O}(J^c).
\label{gen2}
\eeq
Summations
over repeated indices are understood from now on.
Now the  products of
fields transform by action of the co-product of the individual
representations, and thus do span a representation.

Next two  general remarks  are  in order which will be useful below.
First, in the same way as for  ordinary
symmetries  (see Eq.\ref{com0}), one may verify that the transformation law
just recalled
  is consistent with
the assumption that the generators ${\cal O}(J^a)$ and the matrices $\left[
J^a\right]_{m\ell}$
 satisfy the same algebra, which is expressed by the equality
\beq
\left [{\cal O}(J^a), {\cal O}(J^b)\right]= {\cal O}\left (\bigl[J^a,\,
J^b\bigr]\right),
\label{comnaives}
\eeq
which uses the fact that the algebra preserves the coproduct.
However, this is not necessarily true. As we will see below for the case
of $U_q(sl(2))$, consistency of the present co-product action
does not require that the algebra of the generators coincides  with the
one of the matrices: the former may be a suitable extension of the latter
containing additional ``central terms'' that commute with all the $\Psi_\ell$
fields. This will be the subject of section \ref{preamble}. Indeed, we will see
later on  that
it is an algebra of this type  that  will come out from our discussion,
 realized in a somewhat nonstandard way. Thus at this point  we depart from the
general scheme of ref.\cite{MaSc}, where it is assumed that
the generators form a representation of the algebra; we will make further
comments on this below.

Second,
clearly the two sides of Eq.\ref{gen1} are not on the same footing:
one  multiplies by ${\cal O}(J^a)$ on the left, and reads the
transformation law on the right. What would happen if we reverse the
roles of left and right? This brings in the antipode map $S_a^b$ which is
such that (from now on, summation over repeated indices is understood)
$$
\Lambda_{de}^a \> S^{-1}\, ^d_b \> S^{-1}\, ^e_c= \Lambda_{cb}^f
\> S^{-1}\, ^a_f,
$$
\beq
\Lambda_{\ell c}^a \> S^{-1}\, ^\ell_b \> \left [ J^b J^c\right ]_{nm} =
\Lambda_{b \ell }^a \> S^{-1}\, ^\ell_c \> \left [ J^b J^c\right ]_{nm} =
\delta_{nm} \> \epsilon(J^a),
\label{antip}
\eeq
where $\left [ J^b J^c\right ]_{nm}$ is the matrix element in any
representation, and $\epsilon$ is the co-unit. For completeness, let
us recall  that the latter is a complex number such that
\beq
\Lambda_{bc}^a \epsilon(J^c)=\delta_{a, b}
\label{counit}
\eeq
Using the formulae just summarized, it is easy to verify that
Eq.\ref{gen1} is equivalent to
\beq
 \Psi_\ell {\cal O}(J^a)=
\bar\Lambda_{bc}^a {\cal O}(J^b) \Psi_m \left[ J_{(S)}^c\right]_{m\ell}
\label{gen3}
\eeq
where
\beq
J_{(S)}^c=S^{-1}\, ^c_d \> J^d,
\label{gen4}
\eeq
and
\beq
\bar\Lambda_{bc}^a =\Lambda_{cb}^a
\label{lambdabar}
\eeq
is the other  co-product (transpose of the previous one). In our discussion
both possibilities will be useful. We will refer to Eq.\ref{gen1} as
describing the right-action (the generator acts to its right), and to
Eq.\ref{gen3} as describing the left-action.
Thus going from the left-action  to the  right-action
corresponds  to the antipode map. For products of fields, we
obtain an equation similar to Eq.\ref{gen2}:
\beq
 \Psi_{\ell_1} \Psi_{\ell_2}  {\cal O}(J^a)=
\bar\Lambda_{bd}^a {\cal O}(J^b) \Psi_{m_1}\Psi_{m_2}
\bar\Lambda_{ec}^d
\left[J_{(S)}^c\right]_{m_1\ell_1}\left[J_{(S)}^e\right]_{m_2\ell_2},
\label{gen5}
\eeq
 which shows that $\bar \Lambda$  appears as in Eq.\ref{gen3}.

In  ref\cite{MaSc} the general
properties of the generators $\cal O$ were characterized, but no attempt at
an explicit construction was made.
 On the other hand,
the complete study  of the operator algebra of Liouville (2D gravity)
has revealed\cite{G1,G3,CGR2}  that
 a particular
basis of chiral operators noted $\xi_M^{(J)}$  exists   whose  OPA is quantum
group symmetric, with products  of operators transforming by the
coproduct of the individual representations of each.
These $\xi$ should be the operators to which the general
construction just recalled applies.  It is the purpose of the present
paper to show how  this is realized, or to be more precise, how a
suitable redefinition of the formulae just given is directly realized
by the OPA of the $\xi$ fields.
In the present quantum group picture, $M$ is a magnetic quantum number, like
the indices displayed in Eq.\ref{trtheo}, while $J$ is the
spin which characterizes the representation. The OPA of the $\xi$
fields  has been
studied at first for standard representations with $2J$ a positive
integer\cite{G1,G3}, and we discuss this case in
sections \ref{operator realization} and
\ref{central term}. On the other hand, the application of
quantum groups to two dimensional
gravity led us to go away\cite{G1} --\cite{GS3} from this conventional
situation. First one needs
to deal with semi-infinite representations with continuous total spins.
Second there are two dual quantum group $U_q(sl(2))$ and $U_{\qhat}(sl(2))$
such that
$q=\exp(ih)$, $\qhat=\exp(i\hhat)$, and $h\hhat=\pi^2$. For continuous spins,
the complete quantum group structure noted $ U_q(sl(2))\odot  U_{\qhat}(sl(2))$
is
a non trivial combination of the two Hopf algebras. Its overall Hopf algebra
structure
will be derived in section \ref{general}  by making use of the present general
scheme.
The particular case of spin $-1/2$ will be considered in section
\ref{extended},
where it is pointed out that the corresponding $\xi$ fields are required
for the description of the Cartan generator $\Jso$ (with right-action).
Finally, let us note that our results will be weaker than a full realization
of the general scheme of ref.\cite{MaSc} concerning a basic point. Due to the
fact that the algebra of our generators will differ from the standard
$U_q(sl(2))$ one, we cannot yet address the question of the existence
of an invariant vacuum $\vert 0 >$ such that
\beq
{\cal O}(J^a)\vert 0> =\epsilon(J^a)\vert 0>
\label{vac}
\eeq
We will comment further on this point in section \ref{Outlook}.

 \section{Preamble: More general definition of the co-product action for
$U_q(sl(2))$. }
\label{preamble}
Let us now turn to two-dimensional gravity (Liouville theory). There
the enveloping algebra is
$U_q(sl(2))$.  In this latter case, the generators are
$J_\pm$ and $q^{\pm J_3}$ which satisfy
\beq
q^{J_3} J_{\pm} =q^{\pm 1}  J_\pm q^{J_3}, \quad
[J_+, J_-] ={q^{2J_3} -q^{-2J_3}\over q-q^{-1}},\quad
q^{J_3} q^{-J_3}=1,
\label{rel1}
\eeq
and we have the coproduct
\beq
\Lambda(q^{\pm J_3})=q^{\pm J_3}\otimes q^{\pm J_3}, \quad
\Lambda(J_\pm)=J_\pm \otimes q^{J_3}+q^{-J_3}\otimes J_\pm.
\label{coprod}
\eeq
Thus one would write
$$
{\cal O}(q^{\pm J_3}) \Psi_M =
\Psi_N \left[ q^{\pm J_3}\right ]_{NM} {\cal O}(q^{\pm J_3})
$$
\beq
{\cal O}(J_\pm) \Psi_M =
\Psi_N \left[ J_\pm \right ]_{NM} {\cal O}(q^{J_3})+
\Psi_N \left[ q^{- J_3}\right ]_{NM} {\cal O}(J_\pm).
\label{trtheo}
\eeq
We now use upper case indices to agree with later notations.
In the general  scheme of Mack and Schomerus, it is assumed that
the generators acting in the Hilbert space and the matrices of the
transformation law of the fields obey the same algebra. Indeed, it is easy to
verify that
the co-product action just written is compatible with the $U_q(sl(2))$ algebra
for the
generators
\beq
\Jto \Jpmo =q^{\pm 1} \Jpmo \Jto,\quad
\Bigl [\Jpo , \Jmo\Bigr]={\Jto^2-\Jso^2 \over q-q^{-1}}.
\label{naives}
\eeq
However,  this  will
not be true in our construction, which suggests more general alternatives.
 Thus we discuss on a general ground
  the possibility that the matrices satisfy the usual algebra,
while the generators  obey  more general braiding relations. At the
present stage of our understanding, we are not yet able to discuss the
case of a general Hopf algebra. Although we believe that the property we
are discussing is not specific to $U_q(sl(2))$, we specialize to this
case from now on.
Let us  determine the most general algebra
of the operators $\Jpmo$ and $\Jtso$ compatible with the co-product action
Eq.\ref{trtheo}. From Eq.\ref{trtheo} we deduce immediately
$$
 ( q\Jpo \Jto -\Jto \Jpo ) \Psi_M =
\Psi_M ( q\Jpo \Jto -\Jto \Jpo ).
$$
Therefore it follows that
\beq
 q\Jpo \Jto -\Jto \Jpo  = C_+
\label{defcp}
\eeq
where $C_+$ is a central term which commutes with all the
$\Psi$'s. In the same way we derive the analogous relation
\beq
\Jto \Jmo  -q^{-1} \Jmo \Jto = C_-
\label{defcm}
\eeq
Computing the action of $\Bigl[ \Jpo ,\Jmo \Bigr] $ on $\Psi _M $ we get
$$
\Bigl[ \Jpo ,\Jmo \Bigr] \Psi _M = \Psi_N  (q^{-2J_3})_{NM}
\Bigl[ \Jpo ,\Jmo \Bigr] +
$$
$$
\Psi_N [J_+ ,J_- ]_{NM} (\Jto )^2 +
\Psi_N \left ( ( J_- q^{-J_3} )_{NM} C_+ + ( J_+ q^{J_3} )_{NM} C_-
\right )
$$
The last term can be "removed" by introducing the operator $\Do$
which acts as
\beq
\Do \Psi_M = \Psi_N  (q^{-2J_3})_{NM} \Do +
\Psi_N \left ( ( J_- q^{-J_3} )_{NM} C_+ + ( J_+ q^{J_3} )_{NM} C_-
\right )
\label{actd}
\eeq
It is easy to see that it can be taken equal to :
\beq
\Do = \left(C_+\Jmo  +C_-\Jpo \right)\Jso
\label{Dodef}
\eeq
We then get
$$
\left ( \Bigl[ \Jpo ,\Jmo \Bigr] - \Do \right ) \Psi_M =
$$
$$
\Psi_N  (q^{-2J_3})_{NM} \left ( \Bigl[ \Jpo ,\Jmo \Bigr] - \Do
\right )
$$
$$
+\Psi_N [J_+ , J_- ]_{NM} (\Jto )^2
$$
from which it follows that
$$
 \Bigl [\Jpo , \Jmo \Bigr]=
$$
\beq
 {\Jto^2-\Jso^2 \over q-q^{-1}} +\left( C_3 \Jso
+C_+\Jmo  +C_-\Jpo \right)\Jso
\label{defbeta}
\eeq
where $C_3$ is also a central term which commutes  with all
the $\Psi$'s. We note that $\Do$ is defined in fact up to
a term $ (\Jso )^2 $.This freedom is already taken into account
in the term $C_3$ .

At this stage we have derived  the general operator algebra
compatible with the action defined by Eq.\ref{trtheo}. Clearly the commutation
relations Eq.\ref{naives} are recovered for $C_\pm=C_3=0$, and we have found
a  three-parameter extension of $U_q(sl(2))$.
Let us discuss its properties by considering an arbitrary
 representation of this algebra by generators
 denoted $\Jt_\pm$, $q^{\pm \Jt_3}$, satisfying

$$
q\Jt_+ q^{\Jt_3} 	 -q^{\Jt_3}  \Jt_+  = C_+,\qquad
q^{\Jt_3} \Jt_-   -q^{-1} \Jt_- q^{\Jt_3} = C_-
$$
$$[ \Jt_+ ,\Jt_- ] =
{q^{2\Jt_3}-q^{-2\Jt_3} \over q- q^{-1}} + \left( C_3 q^{-\Jt_3} +
 C_+\Jt_- + C_-\Jt_+\right) q^{-\Jt_3}
$$
\beq
 [ C_\pm ,\Jt_a ]= [ C_3 ,\Jt_a] =0.
\label{opeal}
\eeq
At this point the question arises if the extended  algebra Eq.\ref{opeal}
is a Hopf algebra, in particular if we have a (coassociative) coproduct.
This coproduct should be formulated in terms of the generators $\Jt_a$ only,
in contrast to Eq.\ref{trtheo} where both the standard $U_q(sl(2))$ matrices
$[J_a]_{NM}$ and the generators $\Jt_a$ appear. Evidently, this new coproduct
is not of the form Eq.\ref{coprod} as the latter
does not conserve the extended
algebra.

The easiest way to answer this question is to show that, in general,
any representation of this algebra
may be re-expressed in terms of the original algebra
 given by Eq.\ref{rel1} via a linear transformation of the
generators.
 It is immediate to check that the following operators
\beq
 J_\pm =\rho ( \Jt_\pm \pm {C_{\pm} \over{1-q^{\pm 1}}}q^{-\Jt_3} ),\qquad
  q^{\pm J_3} =\rho ^{\pm 1} q^{\pm \Jt_3}
\label{transf}
\eeq
with
$$
 \rho ^{-4} =(q-q^{-1}) (C_+ C_- {{1+q}\over{1-q}} -C_3 ) +1
$$
satisfy the original algebra \footnote{provided that $\rho$ is finite, if
not the algebra can only be recast in a form with $ C_{\pm} =0$,
$C_3 =1/(q-q^{-1})$; we will not discuss this case here .},
 and the new generators still act on the $\Psi_M$ fields according to
Eq.\ref{trtheo} (with
${\cal O}(J_a)$ replaced by the $J_a$ above).
These relations can be inverted very easily
\beq
\Jt_\pm =\rho ^{-1} J_\pm \mp {C_{\pm} \over{1-q^{\pm 1}}} \rho q^{-
J_3},\qquad
 q^{\pm \Jt_3} =\rho ^{\mp 1} q^{\pm J_3}
\label{invtransf}
\eeq
The existence of such a linear transformation
shows that at a formal level the present modification is a
trivial extension of $U_q(sl(2))$. Two remarks are in order about this  point.
First it agrees with what is known for $h=0$ from cohomology
arguments\footnote{We are
indebted to Alvarez-Gaum\'e for mentioning this fact.}.
 Second  this is really true
only  if  we may diagonalize the  central terms $C_\pm$
and work within  an  eigenspace where they may be replaced by
numbers. The situation will be more involved in the coming
field theoretic realization, where  $C_\pm$  are operators
in the Hilbert space of states.

{}From this mapping of the new algebra to the standard one it is
very easy to derive the new coproduct
$$
 \widetilde \Lambda(q^{\Jt_3} ) = \rho ^{\pm 1} q^{\pm \Jt_3} \otimes q^{\pm
\Jt_3}
$$
\beq
\widetilde \Lambda(\Jt_\pm ) = \rho \Jt_\pm \otimes q^{\Jt_3} + \rho ^{-1} q^{-
\Jt_3}
\otimes \Jt_\pm
\pm {C_{\pm} \over{1-q^{\pm 1}}} \rho q^{- \Jt_3} \otimes q^{
\Jt_3}
\label{ncoprod}
\eeq
The action of $\Jpmo $ and $\Jtso$ can be defined with the new
coproduct, the matrix elements appearing being the ones of the new
generators of Eq.\ref{invtransf}.
Using Eq.\ref{transf}, it can be immediately rewritten as
\beq
\widetilde \Lambda(q^{\pm \Jt_3} ) =  q^{\pm J_3}\otimes q^{\pm \Jt_3},\qquad
\widetilde \Lambda(\Jt_\pm ) =  J_\pm \otimes q^{\Jt_3} +  q^{- J_3} \otimes
\Jt_\pm
\label{gcoprod}
\eeq
This shows that both coproducts acting in their own way
on the  $\Psi $'s give in fact the same result. This is quite
important since this shows that our definition of the action
is unambiguous. This action is now compatible with the new group law
in the sense defined by Eq.\ref{opeal}. The new coproduct is
coassociative. Acting on the product of two $\Psi $'s it is easy to see
that the action defined from Eq.\ref{gcoprod} is preserved and that
only matrix elements of the ordinary coproduct appear.
This property is true not only for the generators but also for
all  the operators of the enveloping algebra.

The profound reason is an invariance of the
original coproduct structure constants
$  \Lambda ^{a}_{bc} $ .
There exists a group of matrices $X^a_b$ such that
$$ \Lambda ^{a}_{bc} X^d_a (X^{-1})^c_e =  \Lambda ^{d}_{be} $$
for all b.
Then the generators defined by $ \Jt^a =X^a_b J^a $ satisfy new
commutation relations preserved by a new coproduct  with
structure constants given by
$$
  \widetilde \Lambda ^{a}_{bc} = X^a_d \Lambda ^{d}_{ef} (X^{-1})^e_b
(X^{-1})^f_c
$$
The property of invariance of the $\Lambda  $ implies
$$
  \widetilde \Lambda ^{a}_{bc} =\Lambda ^{a}_{dc} (X^{-1})^d_b
$$
so that
$$
  \widetilde \Lambda ^{a}_{bc} \Jt^b \otimes \Jt^c =\Lambda ^{a}_{bc}
J^b \otimes \Jt^c
$$
In our case of $U_q(sl(2))$ it is easy to show that the general
solution for the matrices X is precisely given by the 3 parameters
relations
of Eqs.\ref{invtransf} and that
 these transformations generate the group $E_2$ .

\paragraph{The particular case where $C_3=1/(q-q^{-1})$:}

In the coming discussion,  $\Jso$ will be on a completely different
footing, and will not be introduced at all in the beginning.
This will be possible since  the generators  will satisfy an algebra which
is
closely related  with the one just written in the  particular case
$C_3=1/(q-q^{-1})$ , where the coefficient of
$\Jso^2$ on the right-hand side of   Eq.\ref{defbeta}
vanishes.  In the present  subsection, we show that  indeed
  $\Jso $
 may be completely eliminated --- assuming that  $C_+C_-$  does not vanish ---
for this particular value\footnote{In fact  we may always reduce the general
situation to
the present one by a
suitable redefinition of the generators. For this value of
$C_3$ , $\rho =\infty$ is equivalent to $C_+ C_- = 0$. }
of $C_3$ from the algebra defined by
Eqs. \ref{defcp}, \ref{defcm}, \ref{defbeta}
if we use the operator $\Do$ introduced in Eq.\ref{Dodef},
$$
\Do=C_+\Jmo \Jso
+C_-\Jpo \Jso .
$$
What happens in practice is that $\Jso $ only appears in the particular
 combination $\Do$.
The derivation goes as follows.
First, using $\Do $ this  algebra may be rewritten as
\beq
 q\Jpo \Jto -\Jto \Jpo  = C_+
\label{defcp1}
\eeq
\beq
\Jto \Jmo  -q^{-1} \Jmo \Jto = C_-
\label{defcm1}
\eeq
\beq
C_+\Jmo
+C_-\Jpo  =\Do  \Jto
\label{D2}
\eeq
\beq
 \Jpo \Jmo -\Jmo \Jpo =\Do +
 {(\Jto )^2 \over q- q^{-1}}.
\label{defbeta1}
\eeq
second, making use of the definition Eq.\ref{Dodef}, one verifies that
 the braiding
relations  of $\Do $ with the other generators are given by
$$
 \Jpmo \Do -q^{\pm 1} \Do \Jpmo =\pm {C_\pm \over q-q^{-1}} \Jto,
$$
\beq
\Jto \Do -q^{\mp 1} \Do \Jto =\pm C_\mp (q-q^{-1}) \Jpmo.
\label{Dcom}
\eeq
The key fact here is that $\Jso $
 does not appear explicitly in the braiding relations just derived.
Finally, substituting  $\Jpmo$ as given by Eqs.\ref{Dcom}
into the previous commutation relations
Eqs.\ref{defcp1} --\ref{defbeta1}, one derives further consistency relations
$$
\Jto \Do ^2 -(q+q^{-1}) \Do \Jto \Do +\Do ^2 \Jto  =
C_+C_-\Jto ,
$$
\beq
\Do \Jto ^2 -(q+q^{-1}) \Jto  \Do \Jto +\Jto ^2 \Do =-C_+C_-(q-q^{-1}),
\label{consist}
\eeq
$$
\Do \Jto ^2 \Do - \Jto \Do ^2 \Jto =
$$
$$
-C_+ C_- (q-q^{-1} ) \Do - C_+ C_- \Jto ^2 .
$$
It is easily seen that this last identity is not independent.
It can be obtained by a combination of the first one multiplied
by $\Jto$ on the right and of the second multiplied by $\Do$
on the left. From this one may verify that the enveloping algebra of
the generators $\Jpmo , \Do , \Jto$ may be entirely derived without ever making
use of
Eq.\ref{Dodef}, so that $\Jso $ has been completely eliminated as
we wanted to show.

Recalling the action of $\Do$ on the
$\Psi _M $ fields,
$$
\Do \Psi_M = \Psi_N (q^{-2J_3})_{NM} \Do +
\Psi_N \left ( ( J_- q^{-J_3} )_{NM} C_+ + ( J_+ q^{J_3} )_{NM} C_-
\right )
$$
we see that we have a coaction of $\Do $ defined by
$$
\widetilde \Lambda ( \Do ) = q^{-2 J_3} \otimes \Do +
( J_+ q^{-J_3} ) \otimes C_- +
( J_- q^{-J_3} ) \otimes C_+
$$

In deriving Eqs.\ref{Dcom},\ref{consist} from Eqs.\ref{defcp1}--\ref{defbeta1}
we have made use of the relation ${\cal O}[q^{J_3}]{\cal O}[q^{-J_3}]=1$.
However,  Eqs.\ref{defcp1} -- \ref{consist} altogether define a consistent
operator algebra
which may be considered on its own, without introducing $\Jso $ at all,
and it is this "weak" version of the original equations which will be realized
by our generators ---in a special way to be described
in section \ref{B+B-alg} below. In fact, our realization
will necessitate a further generalization of
the above considerations in that our generators will depend on position, like
the fields on which they act. Though the derivation leading to
Eqs.\ref{defcp1}--\ref{consist} is not strictly applicable in this situation,
we will demonstrate explicitly that the latter, suitably interpreted, are
realized by our generators.

For later reference, we note here already that a more general
condition could be assumed in place of $\Jto \Jso=1$, namely $\Jto \Jso =C_0$,
where $C_0$ is
central. At the present formal level, this modification is trivial as it
amounts only to a normalization change of $\Jso$, and thus the above arguments
are unchanged (except that we should replace $\Jso \to \Jso C_0^{-1}$ in
Eqs.\ref{Dodef},\ref{defbeta}). However, it will acquire a nontrivial meaning
for the field-theoretic realization discussed below (cf. section
\ref{extended}).
Note also that $\Jto $ and  $\Jso $ do not
play the same role in the original algebra.
In particular the expressions  similar to
Eqs.\ref{defcp} and \ref{defcm} with $\Jto\to  \Jso $ are not
central, but proportional to $\Jso ^2 $.

\section{The Operator realization of $U_q(sl(2))$}
\label{operator realization}
\subsection{The generators}
Our basic tool will be the braiding relations of the
family of chiral primaries $\xi_M^{(J)}$, with $2J$ a positive integer
and
$-J\leq M\leq J$.
 We work on the cylinder $0\leq \sigma \leq 2\pi$,
$-\infty \leq \tau \leq \infty$. Since the $\xi$ fields are only
functions  of $\sigma-i\tau$ (or $\sigma-\tau$ in the Minkowsky case), we
may   restrict ourselves to equal $\tau$ and take it  to be zero.
Thus we work on the unit circle $z=e^{i\sigma}$. It was shown in
ref.\cite{G1} that, for $ \sigma > \sigma' $,
\beq
\xi_M^{(J)}(\sigma) \,\xi_{M'}^{(J')}(\sigma')=
\sum_{-J\leq N\leq J;\> -J'\leq N'\leq J'}\>
(J,J')_{M\, M'}^{N'\, N}\, \xi_{N'}^{(J')}
(\sigma') \,\xi_N^{(J)}(\sigma).
\label{ech}
\eeq
where
$$
(J,J')_{M\, M'}^{N'\, N}=\Bigl(<J,M\vert \otimes <J',M' \vert\Bigr)\>
{\bf R}
\>\Bigl(\vert J,N> \otimes \vert J',N'>\Bigr),
$$
\beq
{\bf R}= e^{(-2ihJ_3 \otimes J_3)}
\bigl(1+ \sum_{n=1}^\infty \,
{\displaystyle{(1-e^{2ih})^{n}\,e^{ihn(n-1)/2} \over
\lfloor n \rfloor \! !}} e^{-ihnJ_3}(J_+)^n \otimes
e^{ihnJ_3}(J_-)^n\bigr).
\label{unvR}
\eeq
For the following, we note
 that Eq.\ref{ech} is valid for $\sigma,\sigma' \in [0, 2\pi]$,
although
in refs.\cite{G1,GS1,GS3} it  is stated to hold for
$\sigma,\sigma' \in [0,\pi]$ only\footnote{This is clear from the
underlying transformation law for hypergeometric functions as recalled
e.g. in  ref.\cite{GN}.}.
We have introduced states noted $|J, M>$ which span the spin $J$
representation of $U_q(sl(2))$ in order to write down the universal R
matrix.  We now show that particular cases of
the formulae just written are very similar to Eq.\ref{trtheo}. There
will be important differences which we will spell out in turn.
In the following, the two Borel subalgebras ${\cal B}_+$ with elements $J_+$
and
$q^{J_3}$, and ${\cal B}_-$ with elements
$J_-$ and $q^{-J_3}$, will be considered
separately, as they will be realized in a different way.

\subsubsection{The  Borel subalgebra ${\cal B}_+$}
Let  $\sigma_+>\sigma$. A particular case of Eq.\ref{ech} is
\beq
\xi_{-{1\over 2}}^{({1\over 2})}(\sigma_+) \xi_{M}^{(J)}(\sigma)=
q^M \xi_{M}^{(J)}(\sigma) \xi_{-{1\over 2}}^{({1\over 2})}(\sigma_+)
\label{brxi3+}
\eeq
\beq
\xi_{{1\over 2}}^{({1\over 2})}(\sigma_+) \xi_{M}^{(J)}(\sigma)=
q^{-M} \xi_{M}^{(J)}(\sigma) \xi_{{1\over 2}}^{({1\over 2})}(\sigma_+) +
$$
$$
{(1-q^2)\over q^{{1\over 2}}} <J, M+1|J_+|J,M>
\xi_{M+1}^{(J)}(\sigma) \xi_{-{1\over 2}}^{({1\over 2})}(\sigma_+).
\label{brxi++}
\eeq
This  exactly coincides with Eq.\ref{trtheo} if we identify, up  to
constants, $\xii_{\demi}(\sigma_+)$ with ${\cal O}(J_+)$, and
$\xii_{-\demi}(\sigma_+)$ with
${\cal O}(q^{J_3})$. The  crucial difference with the general
transformation law is that the role of generators is played by
fields that depend upon the word-sheet variable $\sigma_+$.
This is possible since  the braiding matrix of
$\xii(\sigma_+)$  	 with
a general field $\xi_{M}^{(J)}(\sigma)$ only depends upon the sign
of $(\sigma_+-\sigma)$. Thus we may realize the (${\cal B}_+$ part of)
the transformation Eq.\ref{trtheo} simply by  the $\xii$ fields taken
at an arbitrary point (within the periodicity interval $[0,2\pi]$) such that
this difference is positive.
  Accordingly, we will write, keeping in mind the $\sigma_+$
dependence,
\beq
\Jpop{\sigma_+} \equiv\kappa^{(R+)}_+ \xi_{{1\over 2}}^{({1\over
2})}(\sigma_+), \quad
\Jtop{\sigma_+} \equiv \kappa^{(R+)}_3 \xi_{-{1\over 2}}^{({1\over
2})}(\sigma_+),
\label{Jopdef+}
\eeq
where $\kappa^{(R+)}_+$ and $\kappa^{(R+)}_3$  are normalization constants to
be
specified below.
For later convenience, we add a superscript $R$ to indicate that the
 realization is  by  right-action --- that is by acting to the right.
We then obtain the action by co-product of the form Eq.\ref{trtheo}, that is
\beq
\Jtop{\sigma_+} \xi_{M}^{(J)}(\sigma)=
  \xi_{N}^{(J)}(\sigma)  \left[ q^{J_3}\right ]_{NM} \Jtop{\sigma_+}
\label{act3+}
\eeq
\beq
\Jpop{\sigma_+} \xi_{M}^{(J)}(\sigma)=
 \xi_{N}^{(J)}(\sigma) \left(q^{-J_3}\right)_{NM}\Jpop{\sigma_+} +
\xi_{N}^{(J)}(\sigma) \left[J_+\right]_{NM} \Jtop{\sigma_+},
\label{act+}
\eeq
provided $\kappa^{(R+)}_+$ and $\kappa^{(R+)}_3$ satisfy
\beq
{\kappa^{(R+)}_+\over \kappa^{(R+)}_3}={q^{{1\over 2}}\over 1-q^2}
\label{kappardef}
\eeq
Thus we have derived ${\cal B}_+$ transformations by right-action. The
left-action discussed in the introduction will come out automatically
if we braid starting  from the product $\xi_M^{(J)}(\sigma) \xii_{\pm
\demi}(\sigma_+)$. Since we still have $\sigma_+>\sigma$,
 we use the other braiding matrix. Recall that for $\sigma >\sigma'$ one
has\cite{G1}
\beq
\xi_{M'}^{(J')}(\sigma') \,\xi_{M}^{(J)}(\sigma)=
\sum_{-J\leq N\leq J;\> -J'\leq N'\leq J'}\>
 {\overline {(J',J)}_{M'\, M}^{N\, N'}} \, \xi_{N}^{(J)}
(\sigma) \,\xi_{N'}^{(J')}(\sigma')
\label{echb}
\eeq
where
\beq
{\overline {(J',J)}_{M'\, M}^{N\, N'}}=\left ((J,J')_{M\, M'}^{N'\, N}\right)^*
\label{unvRb}
\eeq
and $*$ means complex conjugate of the matrix elements.
{}From this it is easy to verify that the left-action
 is given by Eq.\ref{gen3} with
$J_{(S)\, +}=-qJ_+$, $ q^{J_{(S)\,3}}=q^{-J_3}$
which are  obtained from ${\cal B}_+$ by the antipode map.
This is expected from the general argument
given in the introduction. It will be convenient to denote   the corresponding
operators by  $\Jpopl{\sigma_+} $, and $\Jsopl{\sigma_+} $.  For a better
readability
of the coming formulae, we always normalize the operators so that the symbols
written  in between the square brackets are  exactly equal to the matrix
which appear in the co-product action. Thus we absorb the proportionality
coefficient
between $J_{(S)\, +}$ and $J_+$ by  changing  the definition of the $\kappa$
coefficients. Altogether, we finally have
\beq
\Jsopl{\sigma_+}=\kappa_3^{(L+)} \xii_{-\demi}(\sigma_+), \quad
\Jpopl{\sigma_+}=\kappa_+^{(L+)} \xii_{\demi}(\sigma_+),
\quad {\kappa_+^{(L+)}\over \kappa_3^{(L+)}}={q^{-\demi}\over 1-q^{-2}}
\label{Jopdefl+}
\eeq
$$
\xi_M^{(J)}(\sigma) \Jsopl{\sigma_+} =\Jsopl{\sigma_+}
\xi_N^{(J)}(\sigma) \left [q^{-J_3}\right]_{NM}
$$
\beq
\xi_M^{(J)}(\sigma) \Jpopl{\sigma_+} =\Jpopl{\sigma_+}
\xi_N^{(J)}(\sigma) \left[ q^{J_3}\right]_{NM}+
\Jsopl{\sigma_+}
\xi_N^{(J)}(\sigma) \left [J_+ \right]_{NM}
\eeq
\subsubsection{The  Borel subalgebra ${\cal B}_-$}
${\cal B}_-$ will be realized by letting the $\xii$ fields act again, provided
we
reverse the role of left and right and make use of the fields
$\xii_{\pm \demi}(\sigma_-)$ with $\sigma_-<\sigma$. Since the discussion
is very similar to the
previous one,  we will be brief. One lets
\beq
\Jmopl{\sigma_-} \equiv\kappa^{(L-)}_- \xi_{-{1\over 2}}^{({1\over
2})}(\sigma_-), \quad
\Jsopl{\sigma_-} \equiv \kappa^{(L-)}_3 \xi_{ {1\over 2}}^{({1\over
2})}(\sigma_-),\quad
{\kappa^{(L-)}_-\over \kappa^{(L-)}_3}={q^{{1\over 2}}\over 1-q^2}.
\label{Jopdefl-}
\eeq
\beq
\Jmop{\sigma_-} \equiv\kappa^{(R-)}_- \xi_{-{1\over 2}}^{({1\over
2})}(\sigma_-), \quad
\Jtop{\sigma_-} \equiv \kappa^{(R-)}_3 \xi_{ {1\over 2}}^{({1\over
2})}(\sigma_-),\quad
{\kappa^{(R-)}_-\over \kappa^{(R-)}_3}={q^{-{1\over 2}}\over 1-q^{-2}}.
\label{Jopdefr-}
\eeq
 The action by co-product is
\beq
\xi_{M}^{(J)}(\sigma) \Jsopl{\sigma_-} =
 \Jsopl{\sigma_-}  \xi_{N}^{(J)}(\sigma)  \left[ q^{-J_3}\right ]_{NM}
\label{act3l-}
\eeq
\beq
\xi_{M}^{(J)}(\sigma)\Jmopl{\sigma_-}=
\Jmopl{\sigma_-}  \xi_{N}^{(J)}(\sigma) \left(q^{J_3}\right)_{NM} +
\Jsopl{\sigma_-}\xi_{N}^{(J)}(\sigma) \left[J_-\right]_{NM}
\label{actl-}
\eeq
\beq
\Jtop{\sigma_-} \xi_{M}^{(J)}(\sigma)  =
  \xi_{N}^{(J)}(\sigma)  \left[ q^{J_3}\right ]_{NM} \Jtop{\sigma_-}
\label{act3-}
\eeq
\beq
\Jmop{\sigma_-} \xi_{M}^{(J)}(\sigma)=
 \xi_{N}^{(J)}(\sigma) \left(q^{-J_3}\right)_{NM} \Jmop{\sigma_-}+
\xi_{N}^{(J)}(\sigma) \left[J_-\right]_{NM}  \Jtop{\sigma_-}.
\label{act-}
\eeq
One sees that ${\cal B}_-$ is generated by left-action, while its antipode is
generated by right-action. This different treatment of the two
Borel subalgebras comes from the fact that for ${\cal B}_+$ (resp. ${\cal
B}_-$),
$q^{-J_3}$ (resp. $q^{J_3}$) does not belong to the algebra, so that
only right- (resp. left-)  action --- where $\Jso $ (resp. $\Jto$) does not
appear ---
may be written.
\subsection{Symmetries of the operator-product algebra}
We have begun to see, and it will be more and more evident in the following,
that  the operators $\Jpop{\sigma_+}$,
$\Jtop{\sigma_+}$,  $\Jmopl{\sigma_-}$,
$\Jsopl{\sigma_-}$,  and so on,  play the same role
as  the generators introduced in ref.\cite{MaSc} and recalled in the
introduction. From now on we will call them  the generators.
 The basic point is that product of $\xi$ fields will satisfy relations
of the type Eq.\ref{gen2} which must be compatible with the
fusion and braiding relations.  Consider a product of two general fields
$\xi_{M_1}^{(J_1)}(\sigma_1)$ $\xi_{M_2}^{(J_2)}(\sigma_2)$. If we
choose $\sigma_+> \sigma_1,\sigma_2 >\sigma_-$, the action of the
generators on each field will be given by the previous analysis, and so
the $\sigma_+$ dependence will be irrelevant. Accordingly, we see that
Eq.\ref{gen2} will apply for ${\cal B}_+$, and we get for $J^a\in {\cal B}_+$,
$$
{\cal O}(J^a)_{\sigma_+} \, \xi_{M_1}^{(J_1)}(\sigma_1)
\xi_{M_2}^{(J_2)}(\sigma_2)=
$$
\beq
\xi_{N_1}^{(J_1)}(\sigma_1)
\xi_{N_2}^{(J_2)}(\sigma_2) \Lambda_{bc}^a\left\{
\Lambda_{de}^b \left[J^d\right]_{N_1M_1}
\left[J^e\right]_{N_2M_2}\right\} {\cal O}(J^c)_{\sigma_+}.
\label{B+prod}
\eeq
Similarly, for $J^a\in {\cal B}_-$ Eq.\ref{gen5} will apply and we get
$$
 \xi_{M_1}^{(J_1)}(\sigma_1)
\xi_{M_2}^{(J_2)}(\sigma_2)\, {\cal O}(J^a)_{\sigma_-} =
$$
\beq
\Lambda_{bd}^a {\cal O}(J^b)  \xi_{N_1}^{(J_1)}(\sigma_1)
\xi_{N_2}^{(J_2)}(\sigma_2)
\Lambda_{ec}^d \left[J_{(S)}^c\right]_{N_1M_1}
\left[J_{(S)}^e\right]_{N_2M_2},
\label{B-prod}
\eeq
 Let us next explicitly verify that  these transformation laws are consistent
with the OPA of the $\xi$ fields. First, in ref.\cite{CGR1}
the complete
fusion of the $\xi$ fields was shown to be given (in the coordinates of the
sphere) by
$$
\xi ^{(J_1)}_{M_1}(\sigma_1)\,\xi^{(J_2)}_{M_2}(\sigma_2) =
 \sum _{J_{12}= \vert J_1 - J_2 \vert} ^{J_1+J_2}
g _{J_1J_2}^{J_{12}} (J_1,M_1;J_2,M_2\vert J_{12})\times
$$
\beq
\sum _{\{\nu\}} \xi ^{(J_{12},\{\nu\})} _{M_1+M_2}(\sigma_2)
<\!\varpi _{J_{12}},\{\nu\} \vert V
^{(J_1)}_{J_2-J_{12}}(e^{i\sigma_1}-e^{i\sigma_2})
\vert \varpi_{J_2}\! >,
\label{fusxi}
\eeq
where $(J_1,M_1;J_2,M_2\vert J_{12})$ are the q-Clebsch-Gordan
coefficients, and $g _{J_1J_2}^{J_{12}}$ are the so-called coupling constants
which
depend on the  spins only. The primary fields $V ^{(J)}_{m}(z)$
whose matrix elements appear on the right-hand side are the so-called
Bloch wave operators, with diagonal monodromy, which are linearly related to
the
$\xi$ fields.  We will come back to them in section \ref{central term}. Let us
next  apply this relation to the two
sides of Eqs.\ref{B+prod} and \ref{B-prod}.
This yields the same consistency condition for both $B_+$ and $B_-$:
$$
\sum_{N_1+N_2=N_{12}}  (J_1,N_1;J_2,N_2\vert J_{12}) \Lambda_{de}^b
\left[J^d\right]_{N_1M_1}
\left[J^e\right]_{N_2M_2}
=
$$
\beq
(J_1,M_1;J_2,M_2\vert J_{12}) \left[J^b\right]_{N_{12}M_{12}},
\label{rec3j}
\eeq
which is just the standard form of the recurrence relation for the $3j$
symbols.
Eq.\ref{rec3j} expresses the fact that the $3j$ symbols realize the
decomposition into
irreducible representations of q tensorial products.
Similarly, we apply Eq.\ref{ech} to
both sides of Eqs.\ref{B+prod} and \ref{B-prod}, assuming for definiteness
that $\sigma_1>\sigma_2$.
One derives   the consistency condition
\beq
(J_1,J_2)_{N_1\, N_2}^{P_2\, P_1} \Lambda_{de}^b \left[J^d\right]_{N_1M_1}
\left[J^e\right]_{N_2M_2}=
\Lambda_{de}^b \left[J^d\right]_{P_2N_2} \left[J^e\right]_{P_1N_1}
(J_1,J_2)_{M_1\, M_2}^{N_2\, N_1}
\label{defunvR}
\eeq

According to Eq.\ref{unvR}, this is equivalent to the condition that the
universal
R matrix interchanges the two coproducts. As is well known, this is true by
definition. A similar conclusion is reached if we choose $\sigma_1<\sigma_2$
instead. It is clear from their derivation that Eqs.\ref{rec3j} and
\ref{defunvR} are consequences of the commutativity of fusion and braiding,
and the Yang-Baxter equation, respectively, and thus of the polynomial
equations.
Eqs.\ref{rec3j} and \ref{defunvR} tell us that given the existence of a
relation
 of the form Eq.\ref{gen1}  in the theory - realized here
by Eqs.\ref{brxi3+},\ref{brxi++} -  the operator algebra of the
$\xi$ fields has to be covariant under the action of the quantum group given by
\beq
\xi_M^{(J)}(\sigma) \to \xi_N^{(J)}(\sigma) \left [J^a\right]_{NM}.
\label{qact}
\eeq
For a product of fields, the statement of covariance becomes
\beq
\xi^{(J_1)}_{M_1}(\sigma_1) \xi^{(J_2)}_{M_2}(\sigma_2) \to
\xi^{(J_1)}_{N_1}(\sigma_1) \xi^{(J_2)}_{N_2}(\sigma_2)
\Lambda_{de}^b \left[J^d\right]_{N_1 M_1}
\left[J^e\right]_{N_2 M_2}
\label{xicoprod2}
\eeq
according to Eq.\ref{B+prod}.
The quantum group action so defined
 coincides with the one introduced in ref\cite{G1} without derivation.

In the present case  the generators are themselves given by the simplest $\xi$
fields
 of spin $\demi$, and we thus find ourselves in a bootstrap situation where we
could try to
derive the braiding and fusion of $\xi$ fields with arbitrary spins
from those of $\xi^{(\demi)}_{\pm \demi}$
and $\xi_M^{(J)}$. In fact, this works for the case where all spins are
half-integer positive, thus multiples of the spins of the generator, as was
shown in refs.\cite{G1},\cite{CGR1},\cite{CGR2} using also the associativity
of the operator algebra.   We remark that on the other hand, the exchange
algebra
in the  more general case of arbitrary continous spins was
derived by an entirely different, direct constructive method in
refs.\cite{GS1},\cite{GS2}
(within the Bloch wave picture ). Notice also
that the considerations above
 apply beyond the level of primaries, as Eq.\ref{fusxi} involves the
contributions of all the descendants as well. Their behaviour is governed by
the general
 Moore-Seiberg formalism\cite{MS}. The last term of Eq.\ref{fusxi} is a number,
and thus not
acted upon when we derive the braiding with the operators ${\cal
O}(J^a)_{\sigma_+}$. This is
consistent with the quantum group
structure since it does not depend on the magnetic quantum numbers $M_i$. Thus
the
quantum group structure  of all the descendants of the $\xi$ fields is the
same. As a matter
of fact, we may use the orthogonality of the 3j symbols to transform
Eq.\ref{fusxi}
into
$$
\sum_{M_1+M_2=M_{12}} (J_1,M_1;J_2,M_2\vert J_{12})
 \xi ^{(J_1)}_{M_1}(\sigma_1)\,\xi^{(J_2)}_{M_2}(\sigma_2) =
$$
\beq
g _{J_1J_2}^{J_{12}}
\sum _{\{\nu\}} \xi ^{(J_{12},\{\nu\})} _{M_1+M_2}(\sigma_2)
<\!\varpi _{J_{12}},\{\nu\} \vert V
^{(J_1)}_{J_2-J_{12}}(e^{i\sigma_1}-e^{i\sigma_2})
\vert \varpi_{J_2}\! >,
\label{fusxi2}
\eeq
Let us denote the left-hand side by
$\xi^{[J_1,J_2](J_{12})}_{M_{12}}(\sigma_1,\sigma_2)$.
It follows from Eqs.\ref{B+prod}, \ref{B-prod}, and \ref{fusxi} that under the
action of the
generators  $\Jpop{\sigma_+}$,
$\Jtop{\sigma_+}$,  $\Jmop{\sigma_-}$,
$\Jsop{\sigma_-}$, with $\sigma_+>\sigma_1,\sigma_2>\sigma_-$, their
transformation laws are
similar to Eqs.\ref{act+}, and \ref{act-} with spin $J_{12}$. Following the
same line as above,
this finally shows that the braiding matrix of this field with any covariant
field of spin,
say $J'$ must obey a relation of the form Eq.\ref{defunvR}, and thus be the
corresponding
universal R matrix. In particular, we see that, since this R matrix is equal to
one if $J_{12}=0$,
$\xi^{[J,J](0)}_{0}(\sigma_1,\sigma_2)$ commutes with any $\xi_M^{(J)}(\sigma)$
field. We
will come back to this important fact below.

\subsection{The algebra of the generators}

 There are two levels which we discuss in turn.
\subsubsection{The algebra within ${\cal B}_\pm$}
Consider first ${\cal B}_+$.
The novel feature of the present generators is that they depend upon
$\sigma$. When we discuss their algebra, we could use the fusion relations to
consider
their  products   at the same point. This is not necessary since the above
quantum group
action depends only on the ordering between the $\sigma$ of the generator
and the $\sigma$ of the covariant field, and it will pay not to do so.
 Thus, when we discuss quadratic relations within ${\cal B}_+$,
 we introduce two points $\sigma_+$ and $\sigma'_+$ both
larger than $\sigma$, and for $J^a, J^b \in {\cal B}_+$,  we have a priori four
  products
${\cal O}(J^a)_{\sigma_+} {\cal O}(J^b)_{\sigma'_+}$,
${\cal O}(J^a)_{\sigma'_+} {\cal O}(J^b)_{\sigma_+}$,
${\cal O}(J^b)_{\sigma_+} {\cal O}(J^a)_{\sigma'_+}$,
${\cal O}(J^b)_{\sigma'_+} {\cal O}(J^a)_{\sigma_+}$. When we let them act, for
instance,
 to the right, it is clear that the right-most will act first irrespective of
the choice of which is at $\sigma$ and which is at $\sigma'$.
On the other hand, for each  choice of
ordering between $\sigma$ and $\sigma'$, it follows from the
braiding relations Eq.\ref{ech} or \ref{echb}  that we have equations  of the
type
\beq
{\cal O}(J^a)_{\sigma_+} {\cal O}(J^b)_{\sigma'_+}=
\rho_{cd}^{ab}
{\cal O}(J^c)_{\sigma'_+} {\cal O}(J^d)_{\sigma_+},
\label{echgen}
\eeq
where $\rho$ is a numerical matrix.
Thus we need only discuss relations between the products
${\cal O}(J^a)_{\sigma_+} {\cal O}(J^b)_{\sigma'_+}$   and
${\cal O}(J^b)_{\sigma_+} {\cal O}(J^a)_{\sigma'_+}$, with $a\not= b$. We will
refer to this particular type of braiding relations as  fixed-point (FP)
commutation relations.
For the specific case we are discussing,
 this means that we have to compare the action of $\Jtop{\sigma_+}
\Jpop{\sigma'_+}$
and $ \Jpop{\sigma_+} \Jtop{\sigma'_+}$. Of  course this amounts to
looking for the operator equivalent of the  matrix commutation relation
\beq
\left[q^{J_3}\right]_{MP}\left[J_+\right]_{PN} = q
\left[J_+\right]_{MP}\left[q^{J_3}\right]_{PN}
\label{B+com}
\eeq
 that holds in any spin $J$ representation. Making use of the explicit form of
the
q Clebsch-Gordan coefficients, one sees that
\beq
 \Jtop{\sigma_+} \Jpop{\sigma_+'} -q
\Jpop{\sigma_+} \Jtop{\sigma'_+} = -q^\demi \kappa^{(R+)}_+\kappa^{(R+)}_3 \>
\xi^{[\demi , \demi](0)}_0(\sigma_+, \sigma'_+),
\label{central+}
\eeq
where $\xi^{[\demi , \demi](0)}_0(\sigma_+, \sigma'_+)$ is the left-hand side
of Eq.\ref{fusxi2}
with $J_1=J_2=\demi$, $J_{12}=M_{12}=0$.
Thus the right hand side does not vanish. However, it follows from the above
discussion
that
\beq
 \xi^{[\demi , \demi](0)}_0(\sigma_+, \sigma'_+) \> \xi_M^{(J)}(\sigma)=
\xi_M^{(J)}(\sigma) \> \xi^{[\demi , \demi](0)}_0(\sigma_+, \sigma'_+)
\label{com+}
\eeq
 Thus our generators  satisfy the ${\cal B}_+$  FP commutation relations up to
a central
term, and we have found the equivalent of Eq.\ref{defcp} of section 2, where
$C_+$ is
replaced by $\xi^{[\demi , \demi](0)}_0(\sigma_+, \sigma'_+)$.

Concerning ${\cal B}_-$ the discussion of the left-action
 is essentially the same as for the right-action realization of ${\cal B}_+$.
One
 finds
\beq
 \Jmopl{\sigma_-} \Jsopl{\sigma_-'} -q
\Jsopl{\sigma_-} \Jmopl{\sigma'_-} = -\kappa^{(L-)}_-\kappa^{(L-)}_3
q^{\demi}\>
\xi^{[\demi , \demi](0)}_0(\sigma_-, \sigma'_-).
\label{centrall-}
\eeq
Note that, since we now act to the left, it is the left-most operator which
acts
first. Thus  on the left-hand side of this equation the ordering is the reverse
of the one
of the matrix relation. On the other hand, section 2 only dealt with
right-action. Making use
of Eqs.\ref{Jopdefl-}, it is easily to write the  following
relation equivalent to Eq.\ref{centrall-}
\beq
 \Jtop{\sigma_-} \Jmop{\sigma_-'} -q^{-1}
\Jmop{\sigma_-} \Jtop{\sigma'_-} = \kappa^{(R-)}_-\kappa^{(R-)}_3 q^{-\demi}\>
\xi^{[\demi , \demi](0)}_0(\sigma_-, \sigma'_-).
\label{central-}
\eeq
This is the operator (FP) realization of Eq.\ref{defcm} of section 2.
\subsubsection{The complete algebra }
\label{B+B-alg}
In order to really compare two successive actions we have to act from the
same side for both, so that we will combine one Borel algebra with the
antipode of the other.
Let us act to the right following section 2. A priori, we have a problem to
extend
the notion of FP commutation relations: so far  the   Borel subalgebras ${\cal
B}_+$
and ${\cal B}_-$ --- as well as   their antipodes  --- are
defined by using  points larger and smaller than $\sigma$, respectively.

Let us open a parenthesis to answer a question which may have come to
the mind of the reader. The fact that we only have two generators
$\Jtop{\sigma_+}$
and $\Jpop{\sigma_+}$ is  of course due to our use of the $\xii$ fields
which have only two components.
A priori, we could start from $\xi_M^{(J)}(\sigma_+)$, with
$-J\leq M\leq J$. This does not help, however,
since their braiding only defines the co-product action of the enveloping
algebra of ${\cal B}_+$, in agreement with the fact that  the $\xi^{(J)}$
fields   may be
obtained by fusion of the $\xii$ fields. At the level of $J=1$, for instance,
one has the correspondence $\xi^{(1)}_{-1}\sim {\cal O} [q^{2J_3}]$ ,
$\xi^{(1)}_{0}\sim {\cal O} [ J_+ q^{J_3} ]$
, $\xi^{(1)}_{1}\sim {\cal O} [ J_+^2 ]$. Thus
the other Borel subalgebra never appears.

\paragraph{Making use of   the monodromy:}
At this point, we will make use of the monodromy properties of the
$\xii$ fields. As already recalled, these fields  are linearly related to Bloch
wave operators with diagonal monodromy --- the explicit formulae will be
recalled in section \ref{central term}.
{}From this, it is straightforward to deduce that
\beqa
\xii_{-\demi}(\sigma+2\pi)&=& \xii_{\demi}(\sigma), \nnn
 \xii_{\demi}(\sigma+2\pi)&=&-q\xii_{-\demi}(\sigma) +2q^{\demi}
\cos(h\varpi) \> \xii_{\demi}(\sigma),
\label{mono+}
\eeqa
where $\varpi$ is the (rescaled) zero-mode momentum of the B\"acklund free
field. Of course, we
also have the inverse relation
\beqa
\xii_{-\demi}(\sigma-2\pi)&=&2q^{-\demi}
\cos(h\varpi) \> \xii_{-\demi}(\sigma) -q^{-1}\xii_{\demi}(\sigma)\nnn
 \xii_{\demi}(\sigma-2\pi)&=&\xii_{-\demi}(\sigma).
\label{mono-}
\eeqa
For the following, it is important to stress that
$\varpi$ is an operator, with
non trivial commutation relations with the $\xii$
fields.
It seems natural to conjecture that if the fields
 $ \xii_{\alpha }(\sigma)$
satisfy braiding relations with $\xi _M^{(J)} $ given
by the R matrix
$(\demi, J)$ (resp. $\overline {(\demi, J)}$ ), then
the translated
fields $\xii_{ \alpha} (\sigma-2\pi)$ (resp. $
\xii_{\alpha}(\sigma+2\pi))$
satisfy braiding relations given by the R matrix
$\overline {(\demi,
J)} $ (resp.$(\demi, J)$ )\footnote{Actually this is an immediate
consequence of the translation invariance of the braiding matrix,
whose position dependence is governed by the step function
$\epsilon(\sigma-\sigma')$. Using the full region of validity
$\sigma,\sigma'\in [0,2\pi]$ of Eq.\ref{ech} according to the remark made
there, the assertion then trivially follows. However we provide an independent
explicit proof below  using only the validity of Eq.\ref{ech} for
$\sigma,\sigma'\in [0,\pi]$. }
 To prove this we need to know how to
commute $\cos(h\varpi) $ with $\xi_M^{(J)} $.
The commutation of the fields $ \xii_{\alpha
}(\sigma)$ is easily derived
from the condition that the translated fields
$\xii_{ \alpha} (\sigma \pm 2\pi)$ have the same
braiding as the
fields $ \xii_{\alpha }(\sigma)$ and the consistency
of the
commutation of $\cos(h\varpi) $ with the translated
fields. One obtains
\footnote{This formula could also be derived directly
from the definition of the $\xi^{\demi}$ fields in
terms
of the $\psi^{\demi}$ fields. It is however not
necessary
to make reference to the $\psi^{\demi}$ fields; at
this stage
we could even forget that $\varpi$ is the (rescaled)
zero mode momentum of the Backlund free field.}

$$
2\cos(h\varpi) \> \xii_{\pm \demi} = 2 q^{\mp 1}
\xii_{\pm \demi}
\cos(h\varpi) \pm q^{\mp \demi} (q-q^{-1})\xii_{\mp
\demi}.
$$
The consistency of the commutation of $\cos(h\varpi)
$ with
 the fusion of two fields $\xi$ leads to
$$
2\cos(h\varpi) \>\xi _M^{(J)} = 2 q^{-2M} \xi
_M^{(J)} \cos(h\varpi)
+ \sum _N \xi _N^{(J)} E_{NM}^{(J)},
$$
where the matrices $E_{NM}^{(J)} $ must satisfy
$$
(J_1,M_1;J_2,M_2\vert J_{12})
E_{N_{12}M_{12}}^{(J_{12})} =
\sum_{N_1 +N_2 =N_{12}} \Bigl[ (J_1,M_1;J_2,N_2\vert J_{12})\times
$$
$$
E_{N_2 M_2}^{(J_2)} \> q^{-2 M_1} \delta _{N_1 M_1 }
+ (J_1,N_1;J_2,M_2\vert J_{12}) E_{N_1 M_1}^{(J_1)}
\>
\delta _{N_2 M_2 }\Bigr].
$$
This defines a recurrence relation which determines all
the
matrix elements $E_{NM}^{(J)} $ from $E_{\pm \demi
\mp
\demi}^{(\demi)}$. Comparing with Eq.\ref{rec3j}, we see that actually
 the matrices $(J_{\pm} q^{-J_3})_{MN}$ satisfy the
same recursion relation. Therefore
$$
E_{NM}^{(J)} = a( J_+ q^{-J_3} )_{NM} +b( J_-
q^{-J_3})_{NM}
$$
The coefficients a and b are determined from the
commutation
with $  \xii_{\pm \demi} $ and we get finally
$$
2\cos(h\varpi) \>\xi _M^{(J)} =
$$
\beq
2  \xi _N^{(J)}
(q^{-2J_3})_{NM} \cos(h\varpi)
+ (q- q^{-1}) \sum _N \xi _N^{(J)}\left ( (J_- -J_+ )
q^{-J_3} \right
)_{NM}
\label{comcos}
\eeq

Eq.\ref{comcos} defines a coproduct action  of
 $ 2\cos(h\varpi) $ on $ \xi _M^{(J)} $ with
$$
\Lambda (2\cos(h\varpi) ) = q^{-2J_3} \otimes 2\cos(h\varpi)
$$
\beq
+(q-q^{-1}) (J_- q^{-J_3} -J_+ q^{-J_3}) \otimes \hbox{Id}
\label{copcos}
\eeq
For the action on two $\xi$ it is sufficient to replace the operators
$q^{-2J_3} $ and $ (J_- q^{-J_3} -J_+ q^{-J_3}) $ by their ordinary
coproduct. Now it is straightforward to prove our conjecture and this allows
us to construct the missing generators $ \Jmop{\sigma_+}$ and
$\Jpop{\sigma_-}$. They are given by
\beq
 \Jmop{\sigma_+} = \kappa_-^{(R+)} \xii_{-\demi}(\sigma _+ -2\pi)
\label{J-+}
\eeq

\beq
\Jpop{\sigma_-} = \kappa_+^{(R-)} \xii_{\demi}(\sigma _- +2\pi)
\label{J+-}
\eeq
with
$$
{ \kappa_-^{(R+)}\over \kappa_3^{(R+)} } =
{ \kappa_-^{(R-)}\over \kappa_3^{(R-)} } =
{ q^{-\demi} \over 1-q^{-2} }
$$
\beq
{ \kappa_+^{(R+)}\over \kappa_3^{(R+)} } =
{ \kappa_+^{(R-)}\over \kappa_3^{(R-)} } =
{ q^{\demi} \over 1-q^{2} }.
\label{kapparatio}
\eeq

The explicit expressions for the monodromy of $\xii_{\pm\demi}(\sigma ) $
lead to the following relation between $\Jpmo ,\Jto $ and $ \cos (h\varpi)$

\beq
2 \cos (h\varpi) \Jtop{\sigma_\pm} = (q -q^{-1})
 (\Jmop{\sigma_\pm} - \Jpop{\sigma_\pm} )
\label{relacos}
\eeq

It is easy to check that the action of $ \cos (h\varpi)$ as defined by
Eq.\ref{relacos} is completely equivalent to the one given
in Eq.\ref{comcos} as it should be.

Finally, we  verify that a FP algebra  comes out  which is a realization of
the algebra Eqs.\ref{defcp1} -- \ref{Dcom} of  section 2.
First, comparing Eq.\ref{defcp} with Eq.\ref{central+} at $\sigma_+
$, and Eq.\ref{defcm} with
Eq.\ref{central-} at $\sigma_+ - 2\pi $, respectively,  we see that we may
identify
\beqa
C_+&=& q^\demi \kappa^{(R+)}_+\kappa^{(R+)}_3 \>
\xi^{[\demi , \demi](0)}_0(\sigma_+, \sigma_+'),\nnn
C_-&=& q^{-\demi} \kappa^{(R-)}_-\kappa^{(R-)}_3 \>
\xi^{[\demi , \demi](0)}_0(\sigma_+ -2\pi, \sigma_+' -2\pi).
\label{Cdef}
\eeqa
For $C_-$, one uses the monodromy properties Eqs.\ref{mono-} to verify that
$$
\xi^{[\demi , \demi](0)}_0(\sigma_+ -2\pi, \sigma'_+ -2\pi)=
\xi^{[\demi , \demi](0)}_0(\sigma_+,  \sigma'_+).
$$
We therefore have (cf. Eq.\ref{kapparatio})
\beq
C_-=-\left({\kappa_3^{(R-)}\over
\kappa_3^{(R+)}}\right)^2 C_+
\eeq
Now we can establish  the FP equivalent of
Eq.\ref{defbeta1},
which takes the form
\beq
 \Jpop{\sigma_+}  \Jmop{\sigma'_+} -\Jmop{\sigma_+}
\Jpop{\sigma'_+} =
\Dop {\sigma_+},{\sigma_+'} +
 {\Jtop{\sigma_+} \Jtop{\sigma'_+}  \over q- q^{-1}}
\label{J+J-sigma}
\eeq
One finds that $\Do $ is given by the remarkable expression
\beq
\Dop {\sigma_+},{\sigma_+'} ={1\over  q-q^{-1}} 2\cos
(h\varpi) C_+(\sigma_+, \sigma_+'),
\label{Ddef}
\eeq
which shows how the generator introduced in section 2
is realized.
Next multiply
both sides of  Eq.\ref{relacos}  by
$C_+(\sigma_+',\sigma_+'')$. This gives the FP
version of Eq.\ref{D2},
\beq
C_+(\sigma_+',\sigma_+'')\Jmop{\sigma_+}
+C_-(\sigma_+',\sigma_+')\Jpop{\sigma_+} =
\Dop{\sigma_+'},{\sigma_+''}
\Jtop{\sigma_+}
\label{D2FP}
\eeq
from which one finally gets
\beq
C_+=-C_-.
\label{Cequal}
\eeq
This leads to the condition
$\left( \kappa_3^{(R+)}\right)^2=\left(
\kappa_3^{(R-)}\right)^2$.
We choose\footnote{
This is of course consistent with the fact that
$q^{-J_3}$ is the
antipode of $q^{J_3}$.}
\beq
 \kappa_3^{(R+)}=\kappa_3^{(R-)}.
\label{kappaeq}
\eeq
The right-action of $\Dop {\sigma_+},{\sigma_+'} $ is found to read
$$
\Dop {\sigma_+},{\sigma_+'} \xi_M^{(J)}(\sigma) =
$$
\beq
\xi_M^{(J)}(\sigma) q^{-2M} \Dop {\sigma_+},{\sigma_+'} +
\xi_N^{(J)}(\sigma) \left [J_--J_+\right]_{NM}
q^{-M} C_+({\sigma_+},{\sigma_+'}),
\label{Dact}
\eeq
as may be derived either from Eq.\ref{J+J-sigma} or from the definition
Eq.\ref{Ddef}. It takes the form of a co-product action, if we define the
coproduct of $D$ by
\beq
\widetilde \Lambda(D)=q^{-2J_3}\otimes D +
( J_- -J_+ )q^{-J_3}\otimes C_+,
\label{Dcoprod}
\eeq

In conclusion we have derived a FP realization of the algebra
Eqs.\ref{defcp1} --\ref{Dcom} of section 2,
where $\Jto $ and $\Jpo $ depend upon one point, while $C_\pm$
and $\Do$ depend upon two points. Clearly this number of points
may be identified  with    a sort of additive grading of the algebra
introduced in section 2, such that
Eqs.\ref{defcp1}, \ref{defcm1}, \ref{defbeta1}  have grading two,
 Eqs.\ref{D2}, \ref{Dcom} grading three, and Eqs.\ref{consist}
gradings five and four respectively. In the present realization,
the equations of grading larger than two are not directly FP
realized, but the following simplified versions are:
In fact we are in the special case where $C_-=-C_+$ are equal
and where $\Dop{\sigma_+},{\sigma_+'} $ is equal to either of them
up to an ($\varpi$ dependent) operator  that does not depend upon the points.
 Thus we may devide both
sides of Eq.\ref{consist}
   by $C_\pm$, which reduces the grading (number of
points) to one. Similarly equations of higher grading are to be divided by
appropriate powers of $C_\pm$, so that in the end the grading is always less
than or equal to two. These simplified relations hold
for the present construction.

Let us summarize our present results. We have obtained two types of
representations for the operators $ J_\pm $,$ q^{J_3} $ in term of
the fields $\xii_\alpha(\sigma_+)$ and $\xii_\alpha(\sigma_-)$
\beq
\begin{array}{ll}
{\textstyle{\Jtop{\sigma_+}} \over
\textstyle{\kappa_3^{(R+)}}} & =\xii_{-\demi}(\sigma_+)
=\xii_{\demi}(\sigma_+ -2\pi) \nnn
{\textstyle{\Jpop{\sigma_+}} \over
\textstyle{\kappa_+^{(R+)}}} & =\xii_{\demi}(\sigma_+)
=q^{\demi} (q^\varpi +q^{-\varpi}) \xii_\demi(\sigma_+ -2\pi)
-q \xii_{-\demi}(\sigma_+ -2\pi) \nnn
{\textstyle{\Jmop{\sigma_+}} \over
\textstyle{\kappa_-^{(R+)}}} & =q^{-\demi} (q^\varpi
+q^{-\varpi}) \xii_{-\demi}(\sigma_+)
-q^{-1} \xii_{\demi}(\sigma_+)
=\xii_{-\demi}(\sigma_+-2\pi)
\end{array}
\label{recap1}
\eeq
The representation for $\sigma_-$ is obtained by the exchange of
$\sigma_+$ and $\sigma_- +2\pi$ ( $\sigma_+ -2\pi$ and $\sigma_-$ ).
We have seen also that $(q^\varpi +q^{-\varpi})$ becomes part
of the enveloping algebra through the relation
\beq
(q^\varpi +q^{-\varpi}) \Jto =(q-q^{-1}) (\Jmo -\Jpo ).
\label{recap2}
\eeq
We remark that one could derive  FP commutation relations not only
for operators at points $ (\sigma_+ ,\sigma_+' )$ or
$ (\sigma_- ,\sigma_-' )$ but also at points
$ (\sigma_+ ,\sigma_-' )$ or
$ (\sigma_- ,\sigma_+' )$. This would introduce new central terms
which commute  with all the $\xi_M^{(J)}(\sigma)$ fields, namely
$$
\xic(\sigma_-+2\pi ,\sigma_+') = \xic(\sigma_-,\sigma_+'-2\pi)
$$
\beq
\xic(\sigma_+, \sigma_-'+2\pi) = \xic(\sigma_+-2\pi,\sigma_-')
\label{recap3}
\eeq

\paragraph{Another viewpoint:}
Could we work in a way that would be more symmetric between  ${\cal B}_+$ and
${\cal B}_-$, without having to use the monodromy to "transport" $\sigma_-$
 to $\sigma_+$ or vice versa?
This is indeed possible by  starting  from an expression
of the form ${\cal O}(J^a)_{\sigma_+}\xi_M^{(J)}(\sigma){\cal
O}(J^b)_{\sigma_-}$,
with $J^a\in B_+$, and $J^b\in B_-$. Then, the relationship  between the two
orderings of the action of $J^a$ and $J^b$ is a consequence of the Yang-Baxter
equation associated with the braiding  operation
$\xii_\alpha(\sigma_+)\xi_M^{(J)}(\sigma) \xii_\beta(\sigma_-)$
$\rightarrow$ $\xii_\gamma(\sigma_-) \xi_N^{(J)}(\sigma)
\xii_\delta(\sigma_+)$. In particular, choosing $\alpha=\gamma=\demi$,
$\beta=\delta=-\demi$ gives back the equation
$<J,M|[J_+, J_-]|J,M>=\lfloor 2M \rfloor$ satisfied by the matrix
representations.
\section{The extended framework and the construction of ${\cal O}[q^{-J_3}]$ }
\label{extended}
In the previous sections, we have exclusively
considered $\xi_M^{(J)}$ fields with half-integer positive spins.
Though we obtained the combination
${\cal O}[D]=(C_+ \Jmo +C_-\Jpo )\Jso $ (with $C_+=-C_-$) it was not possible
within this framework to construct the operator
$\Jso $ itself for right-action (or $\Jto$ for left-action).
The reason for this is in fact easy to understand
already by a classical scale argument:
Consider Eqs.\ref{Dodef} and \ref{defbeta1}.
Since $\Jmo$ and $\Jpo$ are realized by $\xi^{(\demi)}$ fields which
have classical
scale dimension $-1/2$, it follows from Eq.\ref{defbeta1} that ${\cal O}[D]$
has scale dimension $-1$, and this is of course true for our realization
Eq.\ref{Ddef}. On the other hand, Eq.\ref{Dodef} then tells us immediately
that the scale dimension of $\Jso$ must be $+1/2$ , as $C_+$ is realized
by operators with total dimension $-1$.\footnote{Since the coefficient of
$\Jso^2$ is zero in our realization, this causes no conflict with
Eq.\ref{defbeta}.} But this is just the classical dimension
of the $\xi$ fields with spin $-1/2$. Thus we are lead to consider the extended
framework described in refs.\cite{GS1}\cite{GS3} where in particular negative
half-integer spins can be considered. However,
for our purposes here it will be sufficient to
know that their braiding with any $\xi_{M'}^{(J')}$
is still given by Eq.\ref{ech}, now specified to
representations of spin $-\demi$ and $J'$, and that the
leading order fusion of $\xi^{(-\demi)}_{\pm \demi}$
 with $\xi^{(J)}_M$ is also described correctly by the formula for positive
half-integer spins\cite{G1}, so that ($z:=e^{i\sigma}$)
\beq
\xi_{\pm \demi}^{(-\demi)}(\sigma)\xi_{M'}^{(J')}(\sigma')\sim (1-{z'\over
z})^{J'h/\pi}
q^{-\demi M'\mp \demi J'}\sqrt{{ {2J'\choose J'+M'}\over {2J'-1 \choose
J'-\demi+M'\pm \demi}}}
\xi_{\pm \demi+M'}^{(J'-\demi)}(\sigma'),
\label{-1/2fus}
\eeq
The basic observation is now
that $\xi^{(-\demi)}_{\demi}$, which is the formal
inverse of $\xi^{(\demi)}_{-\demi}$ according to Eq.\ref{-1/2fus}, has braiding
relations with the $\xi^{(J)}_M$ which are exactly those appropriate for
$\Jsop{} $, i.e.
\beq
\xi^{(-\demi)}_{\demi}(\sigma_+)\xi^{(J)}_M(\sigma)=q^{-M}\xi^{(J)}_M(\sigma)
\xi^{(-\demi)}_{\demi}(\sigma_+)
\label{-1/2+ech}
\eeq
Similarly, from the $R$-matrix for the other case $\sigma_-<\sigma$ one obtains
\beq
\xi^{(-\demi)}_{-\demi}(\sigma_-)\xi^{(J)}_M(\sigma)=q^{-M}\xi^{(J)}_M(\sigma)
\xi^{(-\demi)}_{-\demi}(\sigma_-)
\label{-1/2-ech}
\eeq
Thus we should identify
\beq
\Jsop{\sigma_+}=\kappa_{-3}^{(R+)}\xi^{(-\demi)}_{\demi}(\sigma_+)
\label{Jsopdef}
\eeq
in analogy with Eq.\ref{Jopdef+},
and
\beq
\Jsop{\sigma_-}=\kappa_{-3}^{(R-)}\xi^{(-\demi)}_{-\demi}(\sigma_-)
\label{Jsopdef'}
\eeq
Analogous  formulae of course describe
the realization of $\Jto$ for left-action. From Eq.\ref{-1/2fus} and the first
of Eqs.\ref{mono+} it follows that
\beq
\xi^{(-\demi)}_\demi(\sigma_+)=\xi^{(-\demi)}_{-\demi}(\sigma_+-2\pi).
\label{mononegspin}
\eeq
Thus we are lead to identify
\beq
\kappa_{-3}^{(R+)}=\kappa_{-3}^{(R-)},
\eeq
similarly to Eq.\ref{kappaeq} of section \ref{B+B-alg}, and analogously for the
left-action case.
Notice however that the coefficients
$\kappa^{(R+)}_{-3}= \kappa^{(R-)}_{-3} $ (resp. $\kappa^{(L+)}_3=
\kappa^{(L-)}_3$ ) are not fixed in terms of $\kappa_+^{(R+)},\kappa^{(R-)}_-$,
 as was the case for their counterparts $\kappa_3^{(R_+)}=\kappa_3^{(R-)}$
 (cf. Eqs.\ref{kappardef},\ref{Jopdefr-}). This is of course a consequence of
the fact that $\Jsop{} $ does not appear in the coproduct action of the other
generators. Since the coproduct Eq.\ref{trtheo}
is asymmetric in $\Jto$ and $\Jso$, the analog of Eq.\ref{defcp} is not valid.
Rather, multiplying it by $\Jso$ from both sides, one would conclude that
\beq
q\Jso \Jpo -\Jpo \Jso =C_+ (\Jso)^2
\label{defcp'}
\eeq
However, in our FP realization this cannot be true since the grading
of the righthand side is 4, while that of the lefthand side is 2.
The proper FP analogue  of Eq.\ref{defcp'} is obtained if - as announced
already at the end of section \ref{preamble} - we now introduce
a new central charge $C_0$, defined by
\beq
C_0(\sigma_+,\sigma'_+)=\Jtop{\sigma_+} \Jsop{\sigma'_+}
\label{defC0}
\eeq
It is obvious that $C_0$ commutes with all the $\xi^{(J)}_M$, and one
can make an expansion of this new central charge around $\sigma_+=\sigma'_+$
as we did for $C_+$ to verify that it is given by a sum of local quantities
with respect to the $\xi$ fields. If we now multiply Eq.\ref{central+}
by $\Jsop{\sigma_+''} $ and $\Jsop{\sigma_+'''} $ from the left and from the
right respectively, we obtain
$$
C_0(\sigma_+'',\sigma_+)\Jpop{\sigma_+'} \Jsop{\sigma_+'''} -q\Jsop{\sigma_+''}
\Jpop{\sigma_+} C_0(\sigma_+',\sigma_+''')
$$
\beq
=-\Jsop{\sigma_+''} C_+(\sigma_+,\sigma_+')
\Jsop{\sigma_+'''}
\label{central+'}
\eeq
Notice that the points $\sigma_+''>\sigma_+>\sigma_+'>\sigma_+'''$ appear
in the same sequence in both terms on the lefthand side of the above equation,
in accord with the fixed point prescription for the case of grading 4.
Similarly, we have in place  of Eq.\ref{central-},
$$
C_0(\sigma_-'',\sigma_-)\Jmop{\sigma_-'} \Jsop{\sigma_-'''} -q\Jsop{\sigma_-''}
\Jmop{\sigma_-} C_0(\sigma_-',\sigma_-''')
$$
\beq
=-\Jsop{\sigma_-''} C_+(\sigma_-,\sigma_-')
\Jsop{\sigma_-'''}
\label{central-'}
\eeq
with $\sigma_-''<\sigma_-<\sigma_-'<\sigma_-'''$.
With $\Jso$ at our disposal, we can now give a concrete sense also to
Eq.\ref{Dodef} within our realization. Indeed, multiplying the FP
equivalent of Eq.\ref{D2}, that is Eq.\ref{D2FP}, by $\Jsop{\sigma_+'''}$
from the right, one obtains
$$
\Dop{\sigma_+'},{\sigma_+''} C_0(\sigma_+,\sigma_+''')
$$
\beq
 =C_+(\sigma_+',\sigma_+'')\Jmop{\sigma_+} \Jsop{\sigma_+'''}
-C_+(\sigma_+',\sigma_+'') \Jpop{\sigma_+} \Jsop{\sigma_+'''}
\label{DodefFP}
\eeq
Thus we now have a FP  realization of the full algebra, with both $\Jto$
and $\Jso $ available as well for right- as for left-action, and
this completes our considerations on the operator realization of the
quantum group action.
\section{Study of the central term}
\label{central term}
The existence of a nontrivial operator which commutes with all the
$\xi_M^{(J)}(\sigma)$ (with a suitable range of $\sigma$) may seem surprising.
Let us therefore investigate the structure of the central term in
Eqs.\ref{central+},\ref{central-} in some more detail. For this purpose, it is
convenient to
reexpress it in terms of Bloch wave operators. The simplest formulae arise by
using the $\psi$ fields  introduced in ref.\cite{G1}. These are related to
the $\xi$ fields as follows
\beq
\xi_M^{(J)}(\sigma) := \sum_{-J\leq m \leq J}\vert J,\varpi)_M^m \>
\psi_m^{(J)}(\sigma),
\label{psidef}
\eeq
$$
\vert J,\varpi)_M^m\,:= \sqrt{\> {\textstyle {2J \choose J+M}}}
\,e^{ihm/2}\,\times
$$
\beq
 \sum_t e^{iht\,(\varpi +m)}
 {\textstyle{J-M \choose (J-M+m-t)/2}}\,
{\textstyle{J+M \choose (J+M+m+t)/2}},
\label{jw}
\eeq
$$
{P \choose Q} := {\lfloor P \rfloor \! !
 \over \lfloor Q \rfloor \! !\lfloor P-Q \rfloor \! !},
\qquad \lfloor n \rfloor \! ! :=
\prod_{r=1}^n \lfloor r \rfloor,
$$
where the variable t takes all values such that the entries of the
binomial coefficients
are non-negative integers. (We consider only the case of half-integer positive
spin here). The symbol $\varpi$ denotes the rescaled
Liouville zero-mode momentum.
Using this relation for $J=\demi$, one finds
$$
\xic(\sigma_+,\sigma_+')=
$$
\beq
2i \left\{\sin [h(\varpi+1)]\psii_{\demi}(\sigma_+)\,
\psii_{-\demi}(\sigma_+')-\sin [h(\varpi-1)]\psii_{-\demi}(\sigma_+)\,
\psii_{\demi}(\sigma_+')\right\} .
\label{centrpsi}
\eeq
In general the index $m$ of the  $\psi$ fields characterizes the shift of
$\varpi$:
\beq
\psi_m^{(J)}(\sigma)\, f(\varpi)= f(\varpi+2m) \,\psi_m^{(J)}(\sigma).
\label{shift}
\eeq
This shows that the central term commutes with $\varpi$.  It then follows from
the (inverse of) Eq.\ref{psidef} that it also commutes with any
 $\psi_M^{(J)}(\sigma)$ field,
if $\sigma_+, \sigma_+' >\sigma$.

How is this possible?
First consider the classical ($h=0$ case). In this limit $h\varpi$ is kept
fixed so that
$$
\xic(\sigma_+, \sigma_+')\propto \psii_{\demi}(\sigma_+)\,
\psii_{-\demi}(\sigma_+')-\psii_{-\demi}(\sigma_+)\,
\psii_{\demi}(\sigma_+').
$$
Classically, $\psii_{\pm \demi}$ are solutions of the Schr\"odinger equation
$(-\partial_\sigma^2+T) \psii_{\pm \demi}(\sigma)=0$. Making use of this fact,
it is
easy to Taylor expand. The first terms read
$$
\psii_{\demi}(x+\epsilon)\,
\psii_{-\demi}(x)-\psii_{-\demi}(x+\epsilon)\,
\psii_{\demi}(x)\sim
$$
$$
 ({\psii_\demi}'(x)\psii_{-\demi}(x)-
{\psii_{-\demi}}'(x)\psii_{\demi}(x))\times
$$
$$
\Bigl\{\epsilon+{\epsilon^3\over 3 !} T(x)+{\epsilon^4\over 4 !}2 T'(x)+
{\epsilon^5\over 5 !}(T''(x)+3T^2(x))+
$$
\beq
{\epsilon^6\over 6 !}( 4T'''(x)+6 T(x) T'(x))+\cdots \Bigr\}
\label{clexp}
\eeq
The first factor is the Wronskian which is a constant, say one. One
sees that the classical central term has an expansion in $\sigma_+-\sigma_+'$
where the
coefficients are polynomials in $T(\sigma_+')$ and its derivatives. Since
 $\psi_m^{(J)}(\sigma)$ is
a primary with weight $\Delta_J$, its Poisson bracket  with $T(\sigma_+')$
reads
$$
\Bigl\{T(\sigma_+'),\, \psi_m^{(J)}(\sigma)\Bigr\}_{\hbox{\footnotesize P.B.}}=
-4\pi \gamma \left(\delta(\sigma-\sigma'_+){\partial \over
\partial\sigma}+\Delta_J\delta'(\sigma-\sigma'_+)\right )
\psi_m^{(J)}(\sigma).
$$
It thus follows that the Poisson bracket of each term of the expansion
Eq.\ref{clexp}
is a sum of derivatives of delta functions which indeed vanishes
 for $\sigma_+, \sigma'_+ >\sigma$
as we wanted to verify.

Let us return to the quantum level. We expect that a similar mechanism will be
at work. Indeed, the operators $\psii$ satisfy a quantum version of the
Schr\"odinger equation.
The expansion Eq.\ref{fusxi} of $\xic$ involves the descendants of unity which
begin
with $T$. One may expect that a general term will  be given by an ordered
polynomial
in the (derivatives of)  $T$. The simplest way to give it a meaning is to order
with respect to the Fourier modes of this operator, in which case the
expectation value between
highest weight states of each term in the expansion will be given by the
expectation
value of  a polynomial of $L_0$. Let us verify this explicitly.
  Using the
differential equation satisfied by $\psii$  fields one may express the
expectation value of the central term as a hypergeometric function.
The overall normalization of the $\psi$ field is derived in ref.\cite{G1}.
Using  formulae given in ref\cite{G1}, \cite{CGR1}, one sees that
$$
<\varpi| \psii_{\pm \demi}(z) \psii_{\mp \demi}(z')|\varpi>=
d_\pm(\varpi)
{z'}^{-\Delta_\demi-\Delta(\varpi)}
{z}^{-\Delta_\demi+\Delta(\varpi)}
$$
\beq
{\left(z'\over z\right)}^{\Delta(\varpi\pm 1)} (1-{z'\over z})^{-h/2\pi}
 F(-{h\over \pi} ,-{h\over \pi}(1\pm \varpi); 1\mp {h\varpi\over \pi} ;{z'\over
z}),
\label{4ptfn}
\eeq
where $F(a,b;c;z)$ is the standard hypergeometric function, and
\beq
d_+=\Gamma(\varpi{h\over \pi}) \Gamma(-(\varpi+1){h\over \pi}),\quad
d_-=\Gamma(-\varpi{h\over \pi}) \Gamma((\varpi-1){h\over \pi}).
\label{ddef}
\eeq
The notation
 $\Delta(\varpi)$ is  such that
 \beq
L_0|\varpi>=\Delta(\varpi)|\varpi>={h\over 4\pi}(\varpi_0^2-\varpi^2)|\varpi>
\label{eigvl0}
\eeq
with $\varpi_0=1+\pi/h$.
The letter $\Gamma$ represents the usual (not q deformed) gamma function.
 The basic relation underlying the fusion
for the present case is the well known relation between hypergeometric
functions
$$F(a,b;c;x)=
{ \Gamma(c)\Gamma(c-b-a) \over \Gamma(c-a) \Gamma (c-b) }
\> F(a,b;a+b-c+1;1-x)+
$$
\beq
 {\Gamma(c)\Gamma(a+b-c) \over \Gamma(a)
\Gamma (b) }\>
(1-x)^{c-a-b} F(c-a,c-b;c-a-b+1;1-x),
\label{relhyp}
\eeq
The particular combination of $\psii$ fields appearing in the central term
is such that  the second term vanishes, and we are left with
$$
<\varpi| \xic(z,z')|\varpi>=
\left({z\over z'}\right)^{h(\varpi-\varpi_0)/2\pi} (z-z')^{1+3h/2\pi}
$$
\beq
{\Gamma(-1-{2h\over \pi})\over \Gamma(-{h\over \pi})}
{\lfloor \varpi+1\rfloor +\lfloor \varpi-1\rfloor
\over \lfloor \varpi\rfloor} F( u_0-u, u_0; 2u_0; 1-{z'\over z})
\label{xic1}
\eeq
where we have let $u=h\varpi/\pi$, $u_0=h\varpi_0/\pi$,
and  $\varpi_0=1+\pi/h$. Making use of one of the standard quadratic
transformation
of Goursat's table and writing explicitly the power series expansion of the
resulting hypergeometric
function one finds (see  appendix):
$$
<\varpi| \xic(z,z')|\varpi>=
(z-z')^{1+3h/2\pi}
{\Gamma(-1-{2h\over \pi})\over \Gamma(-{h\over \pi})}
{\lfloor \varpi+1\rfloor +\lfloor \varpi-1\rfloor
\over \lfloor \varpi\rfloor}  \times
$$
\beq
\sum_{\nu=0}^\infty \left ({ (z-z')^2\over 4zz'}\right)^\nu
({h\over \pi})^\nu \prod_{\ell=1}^\nu
<\varpi|{\Delta(\varpi_{2\ell})-L_0\over (u_0+\demi)_\nu \nu \! !}|\varpi>,
\label{xicoexp}
\eeq
where $\varpi_{2\ell}\equiv \varpi_0+2\ell$ is the momentum
of the highest weight state whose weight is equal to the
conformal weight of the
fields $\psi_m^{(\ell)}$, and $(a)_\nu=\prod_{r=0}^{\nu-1}(a+r)$.
The right hand side is entirely expressed in terms of matrix elements of powers
of $L_0$,
as we had anticipated. Two mathematical properties of the above hypergeometric
function are remarkable, though their meaning is not clear at this time.
First, the form of the arguments is that of a so-called Gauss series
$F(a, b; a+b+\demi; t)$. Hence \footnote{this was discovered by Clausen
in 1828 (!)} its square
is a hypergeometic function of the type  $_3F_2$, and thus the square of the
expectation value Eq.\ref{xic1} again satisfies a linear differential equation,
now of third order. Second, Eq.\ref{xicoexp} becomes particularly simple
at the point 1, corresponding to $\sigma_+-\sigma_+'=\pi$, where it reduces
to a product of $\Gamma$ functions. See again the appendix for details.

\section{The general quantum group structure}
\label{general}
In this section we turn at once to the most general structure.
On the
one hand, we include the dual Hopf algebra $U_\qhat(sl(2))$, with
$\qhat=\exp(i\hhat)$, and $h\hhat=\pi^2$, on the other hand we deal
with the semi-infinite representations  with  continuous spins introduced
in refs. \cite{GS1} \cite{GR1}\cite{GS3}. In this case  the  Hopf algebra
structure
noted $U_q(sl(2))\odot U_\qhat(sl(2))$ is novel, since it cannot be reduced to
a simple
graded tensor product of $U_q(sl(2))$ and $U_\qhat(sl(2))$. Indeed, although
$\xi$  depends upon four quantum numbers for the half-integer case --- that is
  two total spins  $J$, $\Jhat$, and two magnetic numbers  $M$, $\Mhat$ ---
 there are only three independent quantum numbers   for continuous spins:
 the  so-called
effective total spin
$J^e$ and two screening numbers  $N$, $\Nhat$ which are positive integers.
Within the Bloch wave basis of operators with diagonal monodromy, the
corresponding quantum group symbols and their relations to the operator
algebra have been worked out in refs. \cite{GR1},\cite{GS3}. We will not
attempt
here to present the corresponding derivations within the covariant operator
basis (this will be done in ref.\cite{GS4}) but rather concentrate on the Hopf
algebra structure of the extended quantum group, and simply quote formulae
from ref.\cite{GS4} where necessary.

The most convenient
parametrization of the general $\xi$ fields is obtained by writing them as
$\xi_{\Ms \Mshat}^{(\Je)}$,
with  $M^\circ=N-J^e$ and $\Mhat^\circ=\Nhat-\Jhat^e$, with $\Jehat=J^e h/\pi$.
In effect one has two semi-infinite
lowest weight representations since $\Je+\Ms$, and $\Jehat+\Mshat$ are
arbitrary
positive integers.
 The general braiding matrix takes the form\cite{GS4}
\beq
(({\Je}, {\Je}'))_{
\Ms_{\phantom {2}}\! \!\Mshat\! ; {\Ms}' {\Mshat}'}^{\Mstwo \Mstwohat\! ;\,
\Msone \, \Msonehat}=q^{\Je{\Je}'}\qhat^{\Jehat{\Jehat}'}
(\Je,{\Je}' )_{\Ms {\Ms}'}^{\Mstwo \Msone}
(\Jehat,{\Jehat}')_{\Mshat {\Mshat}'}
^{\Mstwohat \Msonehat},
\label{echgenvect}
\eeq
where the two factors are suitable extensions of the universal R matrix of
$U_q(sl(2))$ and
$U_\qhat(sl(2))$ respectively.
This general formula gives, in
particular, the braiding of the spin $\demi$ fields with a general $\xi$ field,
and thus
determines the co-product action in our scheme.
First  consider one of the two Borel subalgebras. Keeping the same definition
of the operators
${\cal O}(J^a)$ for $J^a\in {\cal B}_+$, one gets
$$
\Jtop{\sigma_+}\xi_{M^\circ \Mhat^\circ}^{(J^e )}(\sigma)=
\xi_{M^\circ \Mhat^\circ}^{(J^e )}(\sigma)
e^{ih \Ms} (-1)^{\Nhat} \Jtop{\sigma_+}
$$
$$
\Jpop{\sigma_+}\xi_{M^\circ \Mhat^\circ}^{(J^e )}(\sigma)=
\xi_{M^\circ \Mhat^\circ}^{(J^e )}(\sigma')\,
e^{-ih \Ms}(-1)^{\Nhat}\Jpop{\sigma_+}
$$
\beq
+\xi_{M^\circ+1,\,  \Mhat^\circ}^{(J^e )}(\sigma)\,
\sqrt{\lfloor J^e-M^\circ \rfloor \lfloor J^e+M^\circ+1\rfloor}
(-1)^{ \Nhat}\Jtop{\sigma_+}.
\label{B+gen}
\eeq
This takes the general form Eq.\ref{gen1}, if we introduce the matrix
representation
\beqa
\left[\Ju_+\right]_{\Ps \Pshat;\, \Ms \Mshat}&=&\delta_{\Ps \, \Ms+1}
\sqrt{\lfloor J^e-M^\circ \rfloor \lfloor J^e+M^\circ+1\rfloor}
\delta_{ \Pshat\,  \Mshat}(-1)^{ \Nhat},
 \nnn
\left[\qJu\right]_{\Ps \Pshat;\, \Ms \Mshat}&=&
\delta_{\Ps ;\, \Ms } \delta_{ \Pshat;\, \Mshat}e^{ih \Ms} (-1)^{\Nhat}\equiv
\left[\qJ\right]_{\Ps \Ms} (-1)^{\Nhat}
\label{B+genma}
\eeqa
and keep the same co-product as before. We use underlined letters for the full
generators. This may be rewritten as
\beqa
\left[\Ju_+\right]_{\Ps \Pshat;\, \Ms \Mshat}&=&\left[J_+\right]_{\Ps  \Ms}
\left[(-1)^{\Nhat}\right]_{ \Pshat\,  \Mshat},\nnn
\left[\qJu\right]_{\Ps \Pshat;\, \Ms \Mshat}&=&\left[\qJ\right]_{\Ps \Ms}
\left[(-1)^{\Nhat}\right]_{ \Pshat\,  \Mshat},
\label{genma}
\eeqa
where there appear
the standard matrices $\left[J^a\right]_{\Ps \Ms}$ of the representation of
$U_q(sl(2))$
with spin $\Je$, and the diagonal matrix $(-1)^{\Nhat}$.

Now in addition we have similar definitions where the role of hatted and
unhatted
quantum nubers are exchanged. Letting
\beq
\Jpophat{\sigma_+} \equiv\kappahat_+ \xihat_{{1\over 2}}^{({1\over
2})}(\sigma_+), \quad
\Jtophat{\sigma_+} \equiv \kappahat_3 \xihat_{-{1\over 2}}^{({1\over
2})}(\sigma_+)
\label{Jophdef+}.
\eeq
\beq
{\kappahat_+\over \kappahat_3}={\qhat^{{1\over 2}}\over 1-\qhat^2},
\label{kapparhdef}
\eeq
we get another co-product action
$$
\Jtophat{\sigma_+}\xi_{M^\circ \Mhat^\circ}^{(J^e )}(\sigma)=
\xi_{M^\circ \Mhat^\circ}^{(J^e )}(\sigma)
e^{i\hhat \Mshat} (-1)^{N} \Jtophat{\sigma_+}
$$
$$
\Jpophat{\sigma_+}\xi_{M^\circ \Mhat^\circ}^{(J^e )}(\sigma)=
\xi_{M^\circ \Mhat^\circ}^{(J^e )}(\sigma')
e^{-i\hhat \Mshat}(-1)^{N}\Jpophat{\sigma_+}
$$
\beq
+\xi_{M^\circ,\,  \Mhat^\circ+1}^{(J^e )}(\sigma)
\sqrt{\lfloorhat \Jhat^e-\Mhat^\circ \rfloorhat \lfloorhat
\Jhat^e+\Mhat^\circ+1\rfloorhat}
(-1)^{ N}\Jtophat{\sigma_+},
\label{Bh+gen}
\eeq
where we have introduced the q deformed numbers with parameter $\hhat$
$$
\lfloorhat x \rfloorhat=
\sin (\hhat x)/\sin \hhat
$$
This takes the general form Eq.\ref{gen1}, if we introduce the matrix
representation
\beqa
\left[\Juhat_+\right]_{\Ps \Pshat;\, \Ms \Mshat}&=&\delta_{ \Ps;\,  \Ms}(-1)^{
N}
\sqrt{\lfloorhat \Jhat^e-\Mhat^\circ \rfloorhat \lfloorhat
\Jhat^e+\Mhat^\circ+1\rfloorhat},  \nnn
\left[\qJuhat\right]_{\Ps \Pshat;\, \Ms \Mshat}&=&\delta_{\Ps ;\, \Ms }
\delta_{ \Pshat;\, \Mshat}(-1)^{N} e^{ih \Mshat}.
\label{B+genmah}
\eeqa
In terms of the hatted generators, the co-product takes the same expression
as for the unhatted generators. This may be rewritten as
\beqa
\left[\Juhat_+\right]_{\Ps \Pshat;\, \Ms \Mshat}&=&\left[(-1)^{N}\right]_{
\Ps;\,  \Ms}
\left[\Jhat_+\right]_{\Pshat  \Mshat},
\nnn
\left[\qJuhat\right]_{\Ps \Pshat;\, \Ms \Mshat}&=&\left[(-1)^{N}\right]_{
\Ps;\,  \Ms}
\left[\qJhat\right]_{\Pshat \Mshat},
\label{genmah}
\eeqa
where there appear
the standard matrices $\left[\Jhat^a\right]_{\Pshat \Mshat}$ of
the representation of $U_\qhat(sl(2))$
with spin $\Jehat$. What is the extended $B_+$ algebra?  Clearly each pair
$J_+$, $\qJ$, and
$\Jhat_+$, $\qJhat$ of matrices satisfies the same algebra as before. For the
mixed relations
one trivially gets
\beqa
\Ju_+\Juhat_+&=&\Juhat_+\Ju_+,\quad \qJu \qJuhat=\qJuhat \qJu, \nnn
\Ju_+\qJuhat&=&-\qJuhat \Ju_+,\quad  \Juhat_+\qJu=-\qJu \Juhat_+.
\label{comB+gen}
\eeqa
The case of ${\cal B}_-$ is treated in exactly the same way, and we will be
very brief. One finds
formulae similar to the above with $J_+\to J_-$ and $\qJ\to\qmJ$. It is easy to
see that one still
has the matrix commutation relations
\beq
\Bigl[\Ju_+,\, \Ju_-\Bigr]={(\qJu)^2-(\qmJu)^2\over q-q^{-1}},\quad
\Bigl[\Juhat_+,\, \Juhat_-\Bigr]={(\qJuhat)^2-(\qmJuhat)^2\over
\qhat-\qhat^{-1}}
\label{+-com}
\eeq

Again the above transformations generate Hopf quantum symmetries of the
operator algebra.
In ref.\cite{GS4},
the generalized 3j symbols are shown to be given by
\beq
(\Jge_1, \Musone, \Jge_2, \Mustwo | \Jge_{12})=
(J^e_1, \Msone , J^e_2, \Mstwo |
J^e_{12}+\phat {\pi \over h})
\widehat (\Jhat^e_1, \Msonehat, \Jhat^e_2, \Mstwohat |
\Jhat^e_{12}+p{h\over \pi}\widehat).
\label{fusgen}
\eeq
On the right hand side the factors are  the standard 3j symbols of
$U_q(sl(2))$, and
$U_\qhat(sl(2))$ respectively,  suitably generalized.
We use underlined letters to denote pairs of quantum numbers: $\Musone$ stands
for
the pair
$\Msone$  $\Msonehat$, and so on. For total spins, we use the same convention,
i.e.
$\Ju$ stands for $\Je$ and $\Jehat$. This latter convention is convenient even
though
$\Je$ and $\Jehat$ are not independent ($\Jehat =\Je {h\over \pi}$), since each
of them plays the role of a
total spin, as is clear from the right hand side. The above Clebsch-Gordan
coefficient is non zero
iff
\beq
\Je_1+\Je_2-\Je_{12}=p+\phat{\pi\over h}
\label{TI1}
\eeq
with $p$ and $\phat$ arbitrary positive integers. Using the same argument as in
section 2.2,
one deduces that the above generalized 3j's should satisfy the recurrence
relation
$$
(\Jge_1, \Musone, \Jge_2, \Mustwo | \Jge_{12}) \left [J^a\right ]_{\Pussum,
\Mussum}=
$$
\beq
\sum_{\Pusone+\Pustwo=\Pussum} (\Jge_1, \Pusone, \Jge_2, \Pustwo | \Jge_{12})
\left[\Lambda(J^a)\right]_{\Pusone \Pustwo, \Musone \Mustwo},
\label{recgen}
\eeq
where $\Mussum=\Musone+\Mustwo$.
Using the recurrence relation for the ordinary 3j symbols together with the
fact that
$p$ and $\phat$ are integer, one may verify this relation. Similarly, one may
verify
that the general braiding matrix is the universal R matrix since it satisfies
the appropriate generalization  of Eq.\ref{defunvR}. Thus we have derived the
full Hopf algebra
structure of $U_q(sl(2))\odot U_\qhat(sl(2))$ in the most general case. This
could not be
achieved before despite several attempts to guess the answer\cite{G1,CG1,CGR2}.
We remark that the above structure can be derived independently on the basis
of the extended quantum group symbols alone\cite{GS4}.

\section{Outlook}
\label{Outlook}
Our original observation was that the braiding properties of the
$\xi_M^{(J)}$ fields allow us to  use the $\xii$ fields  as
generators of the quantum group symmetry. This idea led us to
consider  co-product realizations with novel features. In particular
our  generators are position-dependent, and the algebra of the
field transformation laws (here $U_q(sl(2))$) was seen to follow from
 FP commutation relations where only the  quantum
numbers of the generators are exchanged, and not the operators themselves.
This  FP algebra differs from $U_q(sl(2))$, but we showed in general that
the algebra of the field transformation laws and that of the field generators
need not be identical: the latter  may be a suitable extension of the latter
by central operators that commute with all the fields. Our FP algebra
was found to be precisely a realization of such a central extension of
$U_q(sl(2))$.
In establishing this,   we had to overcome the fact that, in the present
approach,
$U_q(sl(2))$ is split into its two natural Borel subalgebras which are most
directly realized by right- and left-actions, respectively. This was possible
using the monodromy properties of the $\xi$ fields, and  thus the Liouville
zero mode momentum $\varpi$ took part in the algebra, which now includes
a new generator which we called $D$.  This last point is rather interesting,
since
so far $\varpi$ did not play any role in the quantum group structure although
its spectrum of eigenvalues  determines the spectrum of Verma modules.
Of course the ultimate aim of the operator realization is to understand how the
Hilbert space is organized  by the quantum group symmetry, and thus how the
generators act on the Verma modules.
The fact  that $\varpi$  appears in some of our generators may  contain a clue
to this problem. There still remain many related questions,
especially the existence of a vacuum state $|0>$ whose general properties
were summarized in Eq.\ref{vac}. To know the invariant vacuum is obviously
important e.g. for the possibility of writing the q-analog of the Wigner-Eckart
theorem, which would provide a very useful tool for the calculation of matrix
elements of covariant operators. On the other hand, it might turn out that
$U_q(sl(2))$ is spontaneously broken, so that no invariant vacuum exists. Let
us make a further general remark in this connection,
namely, matrix realizations of
our centrally extended algebra (Eq.\ref{opeal})  do not have in general highest
(or lowest weight) states in the usual sense. Indeed, since $C_\pm$ and
$q^{\Jt_3}$
commute, we may diagonalize them simultaneously.
A lowest weight state  $|j,c_\pm>$ would have to satisfy
(we use the notation of Eq.\ref{opeal})
$$C_\pm |j,c_\pm>=c_\pm |j,c_\pm>, \quad  \Jt_- |j,c_\pm>=0,\quad
 q^{\Jt_3}|j,c_\pm>=q^j|j,c_\pm>
$$
It then follows from Eqs.\ref{opeal} that
$$
\Bigl[q^{\Jt_3} \Jt_-   -q^{-1} \Jt_- q^{\Jt_3}\Bigr]|j,c_\pm> =0
= c_- |j,c_\pm>,
$$
so that $c_-$ should vanish. Thus, if $c_\pm\not=0$, the representations of the
algebra
have neither lowest nor highest weight states.
Of course, there is always the possiblity to take the short distance limit
of our FP discussion. Then $C_\pm$ tend to zero, and we can have highest or
lowest weight representations (though still a vacuum in the sense of
Eq.\ref{vac} does not exist). However we prefer to consider the general
situation, as we feel that important information
about the symmetry properties may be lost in the limit (cf. below).
The dependence of our generators upon the points reflects the fact that they do
not commute with the Virasoro generators. Thus there exists an interplay
between the
two symmetries, which should play a basic role.

Another striking aspect deserves closer study: we have seen that the central
terms
may be expressed as series of polynomials in the stress-energy tensor (and its
derivatives). Similar series already appeared in the derivation of the infinite
set
of commuting operators associated with the Virasoro algebra\cite{G0}. Thus
there may be
a deep connection between the present centrally  extended $U_q(sl(2))$ algebra
of
our generators and the complete integrability of the Liouville theory.  The
fact that the generators have become dependent  upon another variable (the
position) is reminiscent of the transition from a Lie algebra to a Kac-Moody
algebra.
Thus the full symmetry of the theory may be ultimately much larger than
presently
known. One may hope that
the understanding of this point will allow us to  solve the dynamics
 of the full integrable structure obtained
by including all of the conserved charges.
Clearly, we are still somewhat far from this ideal situation, but one may be
optimistic.

At a more immediate  level, the present scheme may be used to understand the
quantum group
action on the Bloch wave operators $\psi$, whose monodromy is diagonal.
This is interesting since so far, in sharp contrast with the $\xi$ fields,
their
quantum group properties have remained a mystery. In particular, for them
the role of 3j symbols is actually played by 6j symbols. By braiding our
generators
with the Bloch  wave operators, we may define their quantum group
transformations.
In  connection with our previous remarks about highest/lowest weight states, we
may
mention that the change of basis from the $\xi$ fields to the Bloch wave fields
has its counterpart on the generators themselves. This leads to new generators
where  the central extension is
only multiplicative and does not prevent the existence of highest or lowest
weight states.
 This is  described in
another article\cite{cgs2}. Another direction is to consider higher rank
algebras.
Our  construction of   the generators from the defining representation
(and not from the adjoint, as one would expect a priori) used the  fact that,
for
$U_q(sl(2))$ ,  the dimension of the defining representation
 coincides with the dimension of each Borel algebra. For higher  ranks the
counting is completely different, the former being smaller than the latter. Of
course now there  is more than one representation  of lowest dimension.
Finally, it would be interesting to illuminate the connection of the
present analysis with the general framework of Poisson-Lie symmetries
\cite{AT,FL}, and in particular with the dressing symmetries mentioned in the
introduction. It is challenging to find   a unified treatment which includes
both types
of symmetries, explaining how one and the same $U_q(sl(2))$ quantum group
can manifest itself in apparently rather different guises.

\paragraph{Acknowledgements:} We are grateful to   A. Alekseev,
O. Babelon and J. Teschner for
stimulating discussions. One of us (J.-L. G.) is indebted  to
the Theory Division of CERN for financial support while visiting there.

\appendix
\section{More  about the central terms}
\label{appendix 1}
Here we supplement some details on section \ref{central term}. First, we give
the derivation
of Eq.\ref{xicoexp}.  We start from Eq.\ref{xic1} and use the following
quadratic transformation of Goursat's
table (see for instance ref.\cite{EMO}, p.112, equation (26)):
\beq
(1-y)^{a/2}F(a, b; 2b; y)=F({a\over 2}, b-{a\over 2}, b+{1\over 2};
{y^2\over 4(y-1)}),
\label{xic2}
\eeq
which leads to
$$
\left ({z'\over z}\right )^{(u_0-u)/2}F( u_0-u, u_0; 2u_0; 1-{z'\over z})=
F({u_0-u\over 2}, {u_0+u\over 2}, u_0+\demi; {-(z-z')^2\over 4zz'})
$$
Expanding the right hand side, one sees that the expectation value of the
central term between highest-weight states is given by
$$
<\varpi| \xic(z,z')|\varpi>=
(z-z')^{1+3h/2\pi}
{\Gamma(-1-{2h\over \pi})\over \Gamma(-{h\over \pi})}
{\lfloor \varpi+1\rfloor +\lfloor \varpi-1\rfloor
\over \lfloor \varpi\rfloor}  \times
$$
\beq
\sum_{\nu=0}^\infty \left ({ (z-z')^2\over 4zz'}\right)^\nu
\left({h\over \pi}\right)^\nu \prod_{\ell=1}^\nu
{\Delta(\varpi_{2\ell})-\Delta(\varpi)\over (u_0+\demi)_\nu \nu \! !}.
\label{xicexp}
\eeq
Here $(a)_\nu=\prod_{r=0}^{\nu-1}(a+r)$,
and $\varpi_{2\ell}\equiv \varpi_0+2\ell$.  Using the definition \ref{eigvl0}
 we may rewrite the final result under the form
$$
<\varpi| \xic(z,z')|\varpi>=
(z-z')^{1+3h/2\pi}
{\Gamma(-1-{2h\over \pi})\over \Gamma(-{h\over \pi})}
{\lfloor \varpi+1\rfloor +\lfloor \varpi-1\rfloor
\over \lfloor \varpi\rfloor}  \times
$$
\beq
\sum_{\nu=0}^\infty \left ({ (z-z')^2\over 4zz'}\right)^\nu
({h\over \pi})^\nu \prod_{\ell=1}^\nu
<\varpi|{\Delta(\varpi_{2\ell})-L_0\over (u_0+\demi)_\nu \nu \! !}|\varpi>,
\label{xicoexp'}
\eeq
which is seen to agree with Eq.\ref{xicoexp}.

Being a Gauss series, the hypergeometric function in Eq.\ref{xic1}
is known to square to another one of type $_3F_2$. The precise relation is
\beq
\left\{ F({u_0-u\over 2}, {u_0+u\over 2}, u_0+\demi; t)\right\}^2=
 \,_3F_2\left[{u_0-u, u_0+u, u_0 \atop 2u_0, u_0+\demi}, t\right].
\label{myst1}
\eeq
Furthermore,
applying the so-called Watson theorem (see \cite{Slater} p. 54) one deduces
that
 $$
_3F_2\left[{u_0-u, u_0+u, u_0 \atop 2u_0, u_0+\demi}, 1\right]=
\left\{ {\Gamma(\demi) \Gamma\left(\demi+u_0\right)\over
\Gamma\left (\demi(1+u_0+u)\right) \Gamma\left
(\demi(1+u_0-u)\right)}\right\}^2.
$$
Thus we conclude that
\beq
F({u_0-u\over 2}, {u_0+u\over 2}, u_0+\demi; 1)=
{\Gamma(\demi) \Gamma\left(\demi+u_0\right)\over
\Gamma\left (\demi(1+u_0+u)\right) \Gamma\left (\demi(1+u_0-u)\right)}.
\label{myst2}
\eeq
Thus the above expectation value is especially simple at the point $1$. In
terms of the
original variables this corresponds to $\sigma_+-\sigma'_+=\pi$.

\end{document}